\documentclass[reprint,english,twocolumn,amsfonts,amssymb,amsmath]{revtex4-1}

\usepackage[T1]{fontenc}
\usepackage{amsmath}
\usepackage{amssymb}
\usepackage{stackrel}
\usepackage{graphicx}
\usepackage{natbib}
\usepackage{svg}
\usepackage{bbm}
\usepackage{float}
\usepackage{mathtools}
\usepackage[colorlinks=true,linkcolor=blue,citecolor=blue,urlcolor=blue]{hyperref}
\usepackage[nameinlink,capitalise]{cleveref}
\usepackage{multirow}
\usepackage{float}

\makeatletter
\renewcommand{\p@subsection}{\p@section\thesection} 
\renewcommand{\p@subsubsection}{\p@subsection\thesubsection} 
\makeatother

\creflabelformat{equation}{#2\textup{#1}#3}

\newcommand{\ddt}{\frac{\mathrm{d}}{\mathrm{d}t}}
\newcommand{\tightBrack}[1]{\raisebox{0.25ex}{\scalebox{0.8}{#1}}}%
\newcommand*{\br}[1]{\protect\tightBrack[\mathrm{#1}\protect\tightBrack]}
\newcommand{\E}[1]{\langle#1\rangle}
\newcommand{\bigE}[1]{\bigl\langle #1\bigr\rangle}
\newcommand{\exref}[1]{{\color{blue}#1}}

\newcommand{\fgg}{\E{\br{FG}}_{\mathrm{cg}}}


\begin{document}
\title{Emergent structure and dynamics of tropical forest-grassland landscapes}

\author{Bert Wuyts}
\affiliation{College of Engineering, Mathematics and Physical Sciences, 
	University of Exeter, EX4 4QF, UK}
\author{Jan Sieber}
\affiliation{College of Engineering, Mathematics and Physical Sciences, 
	University of Exeter, EX4 4QF, UK}
\email{b.wuyts@ex.ac.uk}
	
\begin{abstract}
	Previous work indicates that tropical forest can exist as an alternative stable state to savanna. Therefore,
	perturbation by climate change or human impact may 
	lead to crossing of a tipping point beyond which there is rapid forest dieback 
	that is not easily reversed. 
	A hypothesised mechanism for such bistability is a feedback between fire and vegetation,
	where fire spreads as a contagion process on grass patches. 
	Theoretical models have largely implemented this mechanism implicitly,
	by assuming a threshold dependence of fire spread on flammable vegetation.
	Here, we show how the nonlinear dynamics and 
	bistability emerge spontaneously, without assuming equations or thresholds 
	for fire spread. We find that the forest geometry causes the nonlinearity 
	that induces bistability. We demonstrate this in three steps. First, we model 
	forest and fire as interacting contagion processes on grass patches, 
	showing that spatial structure emerges due to two counteracting effects 
	on the forest perimeter: forest expansion by dispersal, and forest erosion by fires 
	originating in adjacent grassland.
	Then, we derive a landscape-scale balance equation in which these two effects
	link forest geometry and dynamics: forest expands proportionally to its perimeter, 
	while it shrinks proportionally to its perimeter weighted by adjacent grassland area.
	Finally, we show that these perimeter
	quantities introduce nonlinearity in our balance equation and lead to bistability.
	Relying on the link between structure and dynamics, we propose a forest resilience 
	indicator that could be used for targeted conservation or restoration.
\end{abstract}
\maketitle

\section*{Introduction}
Satellite \cite{Staver2011b,Hirota2011} and ground observations \cite{Dantas2016,Aleman2020}
show that tropical forest (high tree cover) and tropical savanna (low tree cover)
can exist under the same environmental conditions, making the distribution
of tree cover bimodal.
On the one hand, fire exclusion experiments have shown that fire can maintain low tree cover 
\cite{Bond2008}. On the other hand, fire
occurs almost exclusively below a tree cover threshold of about 40\%
\cite{Staver2011b,Hoffmann2012,Archibald2009,Wuyts2017,VanNes2018}, which is
consistent with fire being a contagion process on grass patches \cite{Schertzer2014,Cardoso2022}, while tree patches block fire. Such a highly nonlinear response of fire to grass together with an empirically consistent
response of vegetation to fire was shown to be sufficient for inducing bistability in simple
models \cite{Staver2012}. Taken together, the bimodality, the mutual interaction between fire and
vegetation, and the availability of a plausible underlying mechanism suggest 
that tropical forest and savanna are alternative 
stable states, maintained by a feedback between vegetation and fire \cite{Staver2011b},
and between which transitions would neither be gradual nor easily reversed 
\cite{Lenton2008,Scheffer2015}.  

Bistability of forest and savanna has been studied with a variety of modelling 
approaches, which can be classified as microscopic versus 
mean-field models. The underlying processes concern the spatiotemporal population dynamics of discrete
vegetation patches, which can spread
or block fire. These can be most realistically modelled by microscopic models, such as interacting particle systems \cite{Patterson2021}
or cellular automata \cite{Hebert-Dufresne2018}, which consider the stochastic
dynamics of such discrete constituents interacting in a spatial domain
according to simple rules. However, as microscopic models are hard to analyse,
one usually looks for a coarse-grained approximation that permits analysis.
Mean-field models provide such an approximation, typically in the form of 
a small number of differential equations that describe the average properties of
the considered populations through time, such as cover fractions of each species.
If the averages are taken over the whole landscape, the resulting mean-field model is non-spatial 
and describes macroscopic dynamics via ordinary differential equations (ODEs) \cite{Staver2012,Touboul2018,VanNes2018}. 
If averages are taken over a neighbourhood, the
mean-field model is spatial and describes the dynamics on a mesoscopic scale, 
via partial differential \cite{Wuyts2019,Goel2020}, spatial difference 
(\citealp{Staal2018};~\citealp[spatial mean field in][]{Li2019}), 
or partial integro-differential equations \cite[mean field in][]{Patterson2021}. 
Mean-field models owe their simple closed form to an assumption of statistical 
independence between species' occurrences in space \cite[e.g.][]{Dieckmann2000,Wuyts2022}, which 
permits writing the interaction between any two species as the product of their occurrences. 
However, assuming statistical independence in space implies neglect of spatial structure.

Despite their disregard for spatial structure and resulting biases \cite[e.g.][]{Keeling1999,Dieckmann2000}, 
mean-field models have been
indispensable tools for gaining theoretical insight in 
alternative stable tree cover states in the tropics. 
The Staver-Levin model of tropical tree cover bistability \cite{Staver2012} is a
non-spatial mean-field model in which the variables
represent grass and tree cover fractions 
in the landscape, with interaction between species captured
as the product of their cover fractions. Fire spread is not included explicitly.
Instead, the effects of fire on vegetation are implicitly accounted for by making
the relevant conversion rates a threshold function of grass cover, 
where the threshold corresponds to the point where large contiguous grass patches emerge, also known as
the percolation threshold \cite{Stauffer1994,Christensen2005}.
The Staver-Levin model has provided a first
proof of principle for alternative stable tree cover states in the tropics,
and showed additional complex behaviours, such as cycles and stochastic resonance 
\cite{Staver2012,Touboul2018}.
Spatial mean-field models of the Staver-Levin model further
showed emergent phenomena due to spatial interaction on mesoscopic scales, 
such as travelling and pinning fronts between states (\citealp{Wuyts2017,Li2019,Wuyts2019,Goel2020};
~\citealp[spatial mean field in][]{Patterson2021}), front curvature effects 
\cite{Li2019,Goel2020}, 
pattern formation \cite{Patterson2023}, 
and coexistence 
states \cite{Bastiaansen2022}. Even though they are spatial, they are still 
mean-field models, as they do not consider the fundamental spreading processes
of forest and fire on patches, but use equations, with implicit assumptions on the spatial structure 
of the patches at finer scales than those modelled. 

The effect of this fine-grained spatial structure can only be 
studied via microscopic models. Schertzer \emph{et al}. (2014) \cite{Schertzer2014}
proposed a cellular automaton implementation of the Staver-Levin model
in which the effect of fire is still captured implicitly, as a threshold function of flammable vegetation.
The form of this vegetation-fire relation was obtained from
separate simulations of fire spread as a standard percolation process. 
The cellular automaton and its mean-field approximation were shown to
exhibit bistability. Thereby, Schertzer \emph{et al}.  \cite{Schertzer2014} provided the
first mechanistic explanation of the role of fire
as a percolation process in bistability of tropical tree cover. It
also justifies the qualitative form of the fire-vegetation dependence
assumed in mean-field models. 
The more recent interacting particle system 
by Patterson \emph{et al}. (2021) \cite{Patterson2021} follows 
a similar approach, by implementing fire as a threshold function of neighbourhood grass cover, where
the threshold is assumed to match with that of site percolation \cite{Stauffer1994,Christensen2005}.
However, standard percolation theory assumes that the occurrences of spreading cells
at different points in space are statistically independent (Section 1.1 in \citealp{Stauffer1994}).
Thus, if fire spread is approximated as a standard percolation process,
one disregards the spatial structure of flammable vegetation.
Hence, although the microscopic Staver-Levin models \cite{Schertzer2014,Patterson2021} consider
the fine-grained patch structure, they still rely on a mean-field assumption in their
implicit treatment of fire, making them prone to biases in regimes with spatial structure.
To avoid these biases, microscopic models require explicit consideration of fire spread 
in interaction with the vegetation landscape, such as in 
the cellular automaton by H\'ebert-Dufresne \emph{et al.} \cite{Hebert-Dufresne2018} 
\cite[see also][]{Favier2004}. 
In this cellular automaton, forest bistability emerges only from simple microscopic rules of vegetation 
and fire spread, i.e. without assuming equations or thresholds for the effects of fire. 
Note that larger-scale forest transitions have also been modelled with a cellular automaton, 
with the effects of climate and fire as spatially heterogeneous parameters \cite{Aleman2018}.

\begin{figure*}[t]
	\centering
	\includegraphics[width=1.9\columnwidth]{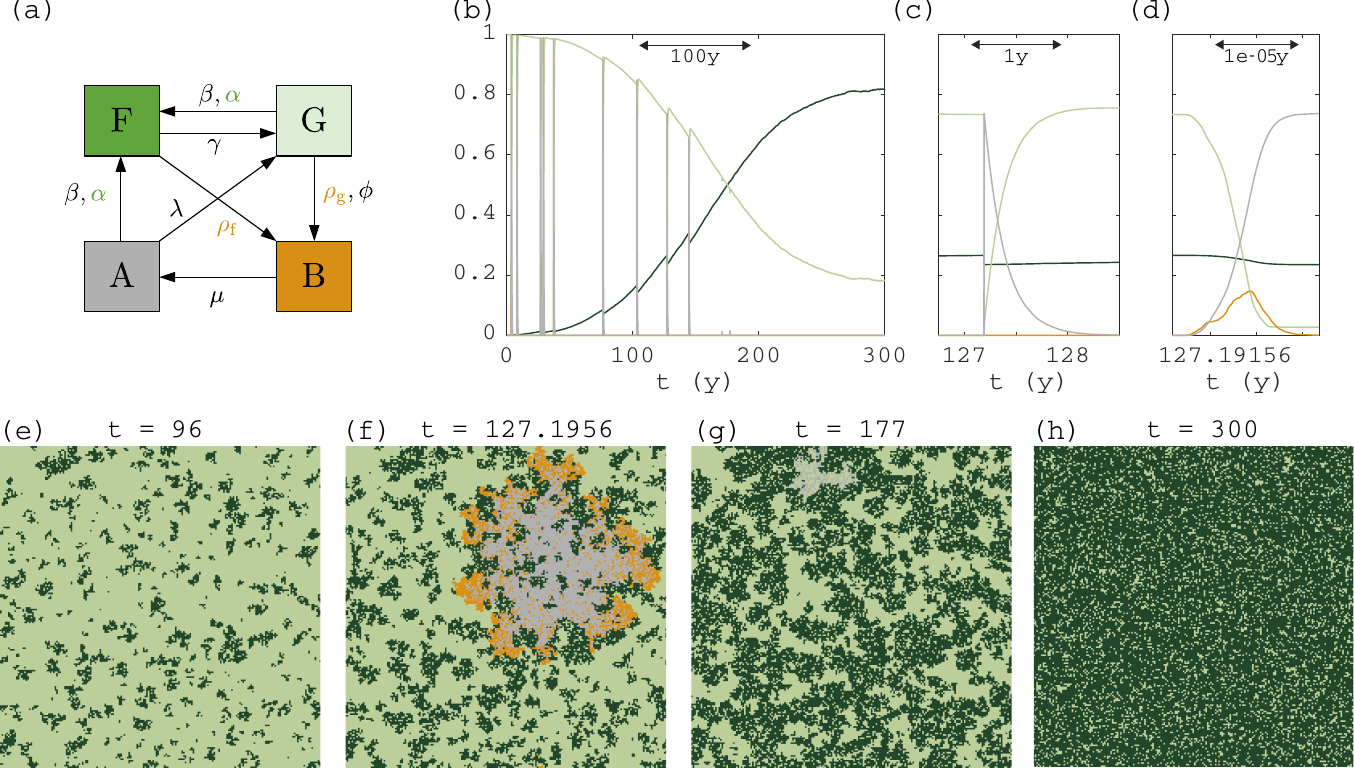}
	\caption{The FGBA stochastic cellular automaton: (a) state transition
		diagram (coloured rates: spread to neighbouring cell, black rates: spontaneous conversion within cell), (b) example time series of a simulation starting at zero
		tree cover, (c,d) $10^2$x and $10^7$x zoom of (b), (e-h) snapshots of
		a simulation at indicated times for low fire ignition rate ($\phi N{=}0.075$). The fire in (f) spreads throughout grassland in the whole domain whereas that in (g) went extinct locally because forest splits grassland in clusters (notice the area of ash near the top). Remaining parameters are shown in \cref{tab:rates}. Domain size: 200x200 cells.}
	\label{fig:FGBAplots}
\end{figure*}

In this work, we examine the spontaneous emergence of nonlinear dynamics and bistability
of tropical forest from the patch-scale rules of forest and fire spread.
We first use
the cellular automaton of H\'ebert-Dufresne \emph{et al.} 
\cite{Hebert-Dufresne2018} to observe
the emergent structure and bistability in simulations.
Next, based on the observations that forest and fire spread occur near the forest perimeter and on separated timescales, 
we set up a macroscopic balance equation of
forest area change (\cref{eq:dFdtclust}). This enables us to analyse the emergent dynamics as a function of the
relevant structure, and will show that the nonlinearity is caused by the forest geometry. Then, we derive a 
forest resilience indicator based on our balance equation, providing a proposed link between the geometry 
and resilience of tropical forest.
Finally, we compare our results against mean-field approximations. This will show that the assumption of
absence of spatial correlations is strongly violated, particularly near the tipping threshold of forest dieback,
while mean-field models still permit accurate expressions for the spatially uncorrelated regime.

\section*{Results}
\subsection*{The FGBA probabilistic cellular automaton}  
The FGBA probabilistic cellular automaton
(adapted from \citep{Hebert-Dufresne2018} -- see \cref{fig:FGBAplots} and Methods) models the 
stochastic dynamics of
tropical vegetation and fire on a square lattice and in continuous time. 
The key empirical facts of tropical forest and fire dynamics captured by the FGBA automaton are: (\emph{i}) fires
only naturally ignite in grasslands but they can spread into forest, (\emph{ii}) fires 
spread more easily in grassland than in forest, such that forests suppress 
fires, albeit imperfectly, (\emph{iii}) forest dynamics occur on a strongly separated
timescale from fire spread and grass regrowth. 

\begin{figure}[t]
	\centering
	\includegraphics[width=1\columnwidth]{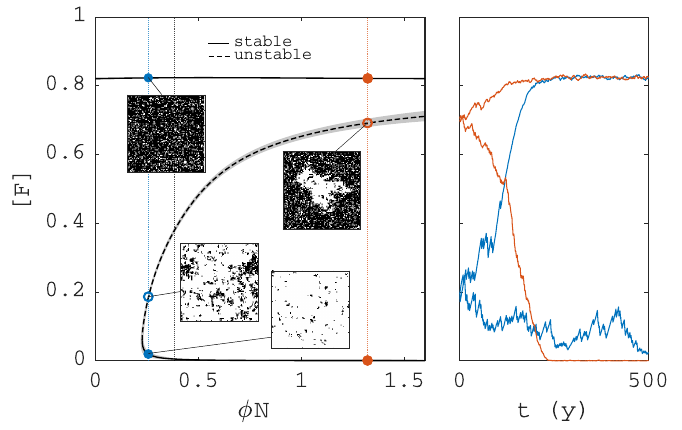}
	\caption{Steady states and bistability of forest area in the FGBA cellular automaton. (a) Bifurcation diagram of forest area fraction $\br{F}$ versus fire ignition rate $\phi$ (shade: two-standard deviation confidence interval of the mean). (b) Simulations initiated at two different points on the saddle ($\phi N{=}0.257$ and $\phi N{=}1.32$). Remaining parameters are shown in \cref{tab:rates}. Domain size: 100x100 cells.}
	\label{fig:intro}
\end{figure}

This results in the following reaction 
rules in the cellular automaton. At any time, each lattice cell
can be in one of four states: $\mathrm{F}$ -- forest, $\mathrm{G}$ -- grass, $\mathrm{B}$ -- burning, $\mathrm{A}$ -- ash. Conversions between these 
states can occur spontaneously or due to spread to neighbouring cells (\cref{fig:FGBAplots}a and \cref{tab:rates}). The spontaneous conversions are: forest recruitment on grass or ash cells due to long-distance seed dispersal or from a homogeneous seed bank ($\mathrm{G}\to\mathrm{F}$ or $\mathrm{A}\to\mathrm{F}$ at rate $\beta$), forest mortality ($\mathrm{F}\to\mathrm{G}$ at rate $\gamma$),
fire ignition on grass cells ($\mathrm{G}\to\mathrm{B}$ at rate $\phi$), and grass regrowth on ash cells 
($\mathrm{A}\to\mathrm{G}$ at rate $\lambda$). The conversions due to spread to neighbours are: forest recruitment due to short-distance seed dispersal on grass or ash cells ($\mathrm{GF}\to\mathrm{FF},\;\mathrm{AF}\to\mathrm{FF}$ at rate $\alpha$), fire spread on grass ($\mathrm{GB}\to\mathrm{BB}$ at rate $\rho_\mathrm{g}$) or on tree cells ($\mathrm{FB}\to\mathrm{BB}$ at rate $\rho_\mathrm{f}$). Chosen parameters are in the ranges empirically justified by 
\cite{Hebert-Dufresne2018} for a square domain of $100$x$100$ cells, with cell size $\Delta x{=}\Delta y{=}30$m. The timescale separation between fire and forest dynamics implies
that the rates satisfy $\rho_\mathrm{g},\rho_\mathrm{f},\mu,\lambda\gg\alpha,\beta,\gamma$. In
particular, we choose
	\begin{align}
		\rho_\mathrm{g},\mu\sim10^6>\rho_\mathrm{f}\sim10^5&\gg1\text{y}^{-1},\label{eq:fast-timescale1}\\
		\alpha,\beta,\gamma\sim10^{[-4,-2]}&\ll\lambda\sim1\text{y}^{-1}.\label{eq:fast-timescale2}
	\end{align}
So, fire spreading and extinction $\rho_\mathrm{g}$, $\rho_\mathrm{f}$, $\mu$ occur on the scale of hours, while grass regrows on ash over months ($\lambda$) and forest spread, growth and mortality $\alpha$, $\beta$, $\gamma$ occur over decades.
We take fire ignition rate $\phi\sim1/N$ such that fires spontaneously occur about once per year in the modelled area.
\Cref{fig:FGBAplots}b--d shows a time profile
for fractions of cells in each state during a simulation with low fire ignition rate $\phi$, starting from
an all-grass state. Due to the low fire ignition rate, the only stable steady state is a nearly closed canopy 
(reached after $300$ years, \cref{fig:FGBAplots}h). Before canopy closure, brief events of rapidly spreading fire counteract a gradual spread of forest. After canopy closure, fires are unable to spread. Timescale separation of forest dynamics (\cref{fig:FGBAplots}b), grass regrowth
(\cref{fig:FGBAplots}c), and fire spread (\cref{fig:FGBAplots}d)
shows clearly. 

\Cref{fig:intro}a shows a bifurcation diagram of steady state forest area in the FGBA cellular automaton, denoted by $\br{F}$ (see \cref{eq:def:xxy}),
versus fire ignition rate $\phi$. Unstable steady states (saddles)
were obtained by applying feedback control to the simulations (see Methods). 
Bistability occurs
above a critical ignition rate $\phi$, with alternative stable states
grassland ($\br{F}{\approx}0$) and forest ($\br{F}{\approx}0.83$). 
Simulations initiated at the saddle will tip randomly up or down 
(\cref{fig:intro}b). Near the lower end of the bistability
range, the saddle solution is fairly homogeneous, but for higher $\phi$ 
values, a single hole of grass in forest arises (insets in \cref{fig:intro}a).

\subsection*{Fast and slow subprocesses} 
The timescale separation (\cref{eq:fast-timescale1,eq:fast-timescale2}) permits treatment of the joint vegetation and fire dynamics
as a fast-slow system. Fire spread occurs on the fast timescale, where
the vegetation landscape is treated as constant. Forest dynamics occur on the slow
timescale, where the effects of fire are a steady state function of the vegetation 
landscape.
\paragraph{Fast process: fires spreading in a given landscape}
On the timescale of a single fire event, forest dynamics are negligible 
($\alpha,\beta,\gamma\ll1/$day)
such that we can consider the total landscape
of forest patches as fixed. For each ignition event, this results in the following 
dynamics. A fire ignites on a grass cell, then spreads
across its grassland cluster at a rate $\rho_\mathrm{g}$ per $\mathrm{BG}$ pair, after which it reaches 
the interface with adjacent forest, where it starts intruding the forest at a rate
$\rho_\mathrm{f}$ per $\mathrm{BF}$ pair. At any time, a burning cell can stop burning spontaneously, 
converting to ash at a rate $\mu$. The probabilities of fire spreading into a neighbouring grass or forest cell
before the originating cell stops burning are given by
	\begin{equation}\label{eq:occupation:prob}
		p_\mathrm{g}:=\frac{\rho_\mathrm{g}}{\rho_\mathrm{g}+\mu}=0.9, \qquad p_\mathrm{f}:=\frac{\rho_\mathrm{f}}{\rho_\mathrm{f}+\mu}=0.1,
	\end{equation}
where we have shown the chosen values in our simulations  
\cite[adopted from][]{Hebert-Dufresne2018}. Since regrowth of grass and ignition of new fires 
occur at a much slower rate than fire spread ($\phi N\lesssim\lambda\ll\rho_\mathrm{f},\mu,\rho_\mathrm{g}$) and our domain is relatively small (see \exref{Section~S2B}
), 
we observe repeated fire spreading events well separated in time (\cref{fig:FGBAplots}b), 
each ending with spontaneous extinction.

When a fire in grassland cluster with index $j$ reaches its interface with adjacent forest,
the resulting forest loss due to this single fire event can be approximated as
(see Methods):
	\begin{equation}\label{eq:forest:loss}
		\Delta_{\mathrm{F},j}^{\mathrm{loss}}:=p_\mathrm{f}\br{FG}_{j},
	\end{equation}
where $\br{FG}_{j}$ counts the number of forest cells adjacent to grassland cluster $j$ 
(with both sides of the equation optionally normalised by $N$). This approximation relies on the assumptions that the fire reaches the 
whole interface with forest (i.e. $p_\mathrm{g}\to1$) and only once per fire (i.e. 
$\rho_\mathrm{g}\gg\lambda\gg\phi N$), and that $p_\mathrm{f}$ is small.

\paragraph{Slow processes: forest demography and fire damage}
Forest demography and loss due to repeated fires occur on the slow timescale.
Writing the number of forest-grass neighbour pairs as $\br{FG}$ (divided by $N$, 
equivalently the total perimeter of forest or grass patches, see \cref{eq:def:xxy}), 
the dynamics for tree recruitment
and mortality result in an expected rate of change for $\br{F}$:
	\begin{equation}
		\label{eq:slow:gain}
		\Delta_{\mathrm{F}}^{\mathrm{gain}}:=\beta\br{G}-\gamma\br{F}+\alpha\br{FG}.
	\end{equation}
In \cref{eq:slow:gain}, the rates of change are $\beta\br{G}$ for spontaneous forest growth on grass, $\gamma\br{F}$ for spontaneous forest mortality, and $\alpha\br{FG}$ for spread of forest into grass at its perimeter. 

The rate of forest erosion at its perimeter due to fire damage over many fire events 
is the weighted sum over all grass clusters $j{=}1,...,n_\mathrm{c}$, i.e.
	\begin{align}
		\Delta_{\mathrm{F}}^{\mathrm{loss}}&:=\sum_{j=1}^{n_\mathrm{c}}\phi N\br{G}_j\Delta_{\mathrm{F},j}^{\mathrm{loss}}
		=\phi N p_\mathrm{f}\sum_{j=1}^{n_\mathrm{c}}\br{G}_j\br{FG}_{j},\label{eq:slow:loss}
	\end{align}
where $\br{G}_j$ is the fraction of $\mathrm{G}$ cells in grass cluster
$j$ (so, $\br{G}{=}\sum_{j=1}^{n_\mathrm{c}}\br{G}_j$), $\phi N\br{G}_j$ is the rate at which fires spontaneously ignite in grass cluster $j$ ($\phi$ is the rate per cell and $N\br{G}_j$ is the area of the cluster), and $\Delta_{\mathrm{F},j}^{\mathrm{loss}}{=}p_\mathrm{f}\br{FG}_j$ is the conversion of forest to ash caused by each fire event (see \cref{eq:forest:loss}) (note also that $\br{FG}{=}\sum_{j=1}^{n_\mathrm{c}}\br{FG}_j$). By defining the grassland-weighted forest perimeter as
	\begin{equation}\label{eq:defFGavcg}
		\fgg:=\sum_{j}^{n_\mathrm{c}}\frac{\br{G}_j}{\br{G}} \br{FG}_j,
	\end{equation}
the expression for forest loss 
becomes
	\begin{equation}
		\label{eq:slow:loss2}
		\Delta_{\mathrm{F}}^{\mathrm{loss}}=\phi\, p_\mathrm{f}\, N\br{G}\fgg.
	\end{equation}
The grassland-weighted forest perimeter $\fgg$ is the average perimeter
of forest clusters weighted by the relative size of their adjacent grass cluster.

\begin{figure}[t]
	\centering
	\includegraphics[width=1\columnwidth]{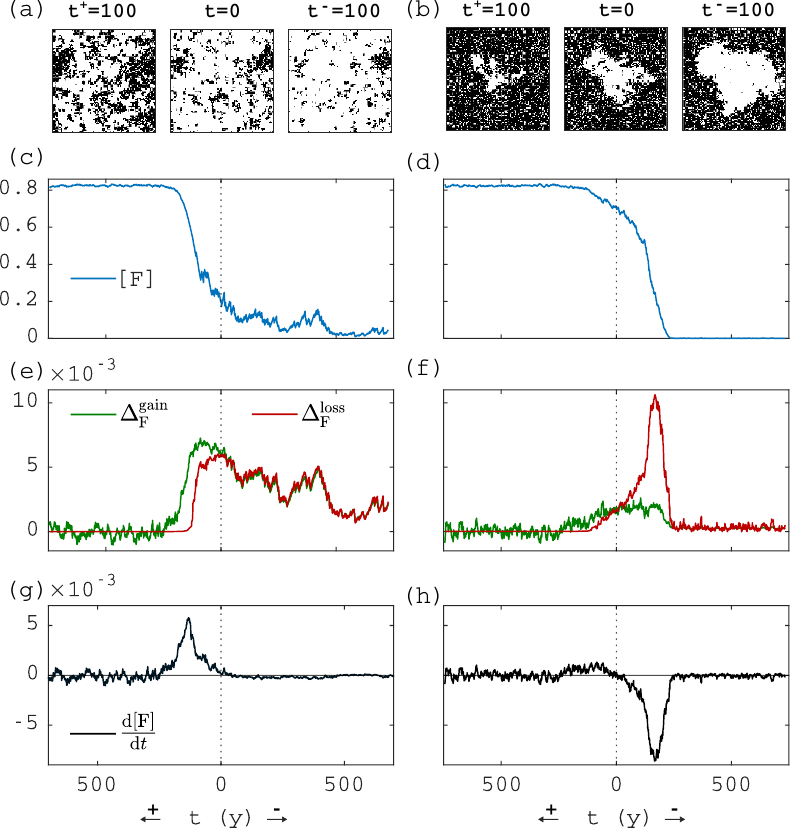}
	\caption{Rate of change according to \cref{eq:dFdtclust}. (a,b) Spatial snapshots at indicated times, (c,d) time series of $\br{F}$, (e,f) time series of $\Delta_{\mathrm{F}}^{\mathrm{gain}}$, $\Delta_{\mathrm{F}}^{\mathrm{loss}}$, (g,h) time series of the right-hand side of \cref{eq:dFdtclust} (gain minus loss). At $t{=}0$, the simulation is started on the saddle (on the left, $\br{F}(0)\approx0.2$ and on the right, $\br{F}(0)\approx0.7$, see blue and red circles in \cref{fig:intro}). Towards the left (along $t^+$), a simulation that tips up and towards the right (along $t^-$) a simulation that tips down is shown. Parameters are shown in \cref{tab:rates}. Columns correspond to leftmost and rightmost vertical dashed lines in \cref{fig:intro} ($\phi N{=}0.257$ and $\phi N{=}1.32$). Domain size: 100x100 cells.}
	\label{fig:gainlossterms}
\end{figure}

\subsection*{Emergent slow dynamics} 
We now form the balance between the slow processes discussed above,
assuming fire converts trees immediately to grass (i.e. $\lambda\gg\phi N$).
The resulting expected rate of forest area change during a short time interval is
	\begin{align}
		\langle\frac{\mathrm{d}\br{F}}{\mathrm{d} t}\rangle =&     \Delta_{\mathrm{F}}^{\mathrm{gain}}-\Delta_{\mathrm{F}}^{\mathrm{loss}},\nonumber\\
		\frac{\mathrm{d}\br{F}}{\mathrm{d}t}=&\beta \br{G} - \gamma \br{F} + \alpha \br{FG} - \phi p_\mathrm{f} N\br{G}\fgg,\label{eq:dFdtclust}
	\end{align}
where we used \cref{eq:slow:gain,eq:slow:loss2}, and assumed on the left-hand side that $N$ 
is sufficiently large, such that, via the law of large numbers, 
$\langle \mathrm{d}\br{F}/\mathrm{d}t\rangle\approx \mathrm{d}\br{F}/\mathrm{d}t$.
\Cref{eq:dFdtclust} can be understood intuitively as 
forest and grass competing for space within clusters (spontaneous terms) 
and at their interface (interaction terms). A larger interface $\br{FG}$ leads 
simultaneously to faster forest spread (proportional to its perimeter $\br{FG}$), and to increased exposure to fires (proportional to its grassland-weighted perimeter $\fgg$). Fires are most
damaging to forest when $\br{G}$ forms a single cluster, i.e. $\fgg{=}\br{FG}$, 
such that each fire reaches the whole interface. Conversely, when forest patches break $\br{G}$
into several clusters $\fgg$ is smaller than $\br{FG}$, such that several ignitions are required to have
the same effect, slowing forest erosion down. Additionally, the total amount of grass $N\br{G}$ determines the number of ignitions
and hence the rate at which grass spreads into forest. The parameters determine the relative weight
of each of the discussed effects. 

\Cref{fig:gainlossterms}
shows example simulations along trajectories starting from the saddle equilibria of \cref{fig:intro}, showing
forest area $\br{F}$ in space and time (a--d), the gain/loss terms $\Delta_{\mathrm{F}}^{\mathrm{gain}}$ and $\Delta_{\mathrm{F}}^{\mathrm{loss}}$ defined in \cref{eq:slow:gain,eq:slow:loss2} (e--f), and the right-hand side of \cref{eq:dFdtclust}
(gain minus loss, g--h). The left column of \cref{fig:gainlossterms} shows simulations for
low fire ignition rate $\phi$ and low $\br{F}(0)$, and the right column for high $\phi$ and high $\br{F}(0)$. Each 
column shows two realisations, both starting from the same saddle steady state. One 
realisation evolves toward high forest cover, 
shown on axis $t^+$ (increasing to the left from $t{=}0$), the other realisation evolves toward low forest cover,
shown on axis $t^-$ (increasing to the right from $t{=}0$). In the stable steady states, gain (green) and loss (red)
terms vary around the same mean. On the saddle (at $t{=}0$), 
gain and loss functions cross, indicating that the steady states
and changes are accurately captured by \cref{eq:dFdtclust}. The largest changes in forest cover $\br{F}$ occur when there are large changes in forest loss due to fire. 
The snapshots in 
\cref{fig:gainlossterms}a,b show that the high-cover state changes as
an expanding/contracting hole in the forest, whereas the low-cover state
is more homogeneous. 

\subsection*{Emergent nonlinear relations} 
\begin{figure}[t]
	\centering
	\includegraphics[width=1.\columnwidth]{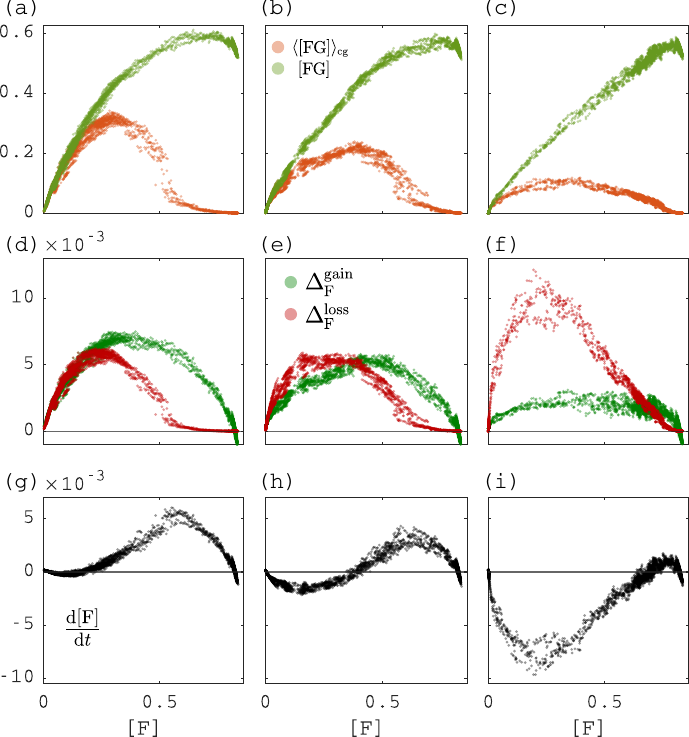}
	\caption{Emergent relations between perimeter quantities and forest area $\br{F}$: (a--c) forest perimeter $\br{FG}$ and grassland-weighted forest perimeter $\fgg$, (d--f) forest gain terms and loss terms in \cref{eq:slow:gain,eq:slow:loss2}, (g--i) forest area rate of change $(\mathrm{d}/\mathrm{d}t)\br{F}$ from \cref{eq:dFdtclust}. Columns correspond to vertical dashed lines in \cref{fig:intro} ($\phi N{=}0.257,\phi N{=}0.38,\phi N{=}1.32$). Domain size: 100x100 cells.}
	\label{fig:scatterplots_changes}
\end{figure}

\Cref{eq:dFdtclust} explains how the rate of change of $\br{F}$ depends on the perimeter quantities $\br{FG}$ and $\fgg$.
\Cref{fig:scatterplots_changes}a--c
shows a scatterplot of $\br{FG}(t)$ and $\fgg(t)$ versus $\br{F}(t)$ for three
different values of $\phi$, and for an ensemble of simulations starting from the saddle in \cref{fig:intro}a, with each point being a value 
observed at a discrete observation time. Remarkably, we observe that
$\br{FG}$ and $\fgg$ lie on a narrow band around some steady state 
functions $\br{FG}^*$ and $\fgg^*$ of $\br{F}$ (and $\phi$), which 
implies that $\br{FG},\fgg$ are changing on
a much faster timescale, making them slaved to $\br{F}$.
\Cref{fig:scatterplots_changes}d--f shows the terms on the right-hand side of
\cref{eq:dFdtclust} depending on $\br{F}$, splitting between gain and loss
terms $\Delta_{\mathrm{F}}^\mathrm{gain},\Delta_{\mathrm{F}}^\mathrm{loss}$, as defined in \cref{eq:slow:gain,eq:slow:loss2}. Steady states occur when
gain equals loss ($\Delta_{\mathrm{F}}^\mathrm{gain}{=}\Delta_{\mathrm{F}}^\mathrm{loss}$). The resulting plot of $\mathrm{d}\br{F}/\mathrm{d}t$
versus $\br{F}$ in \cref{fig:scatterplots_changes}g--i clearly shows the bistability of
$\br{F}$.

Replacing the quantities $\br{FG}$ and $\fgg$ by their steady-state functions $\br{FG}^*$ and $\fgg^*$ results in the single-variable ODE for $\br{F}$,
	\begin{equation}\label{eq:dFdtclust2ss}
		\frac{\mathrm{d}\br{F}}{\mathrm{d}t} = \beta\br{G} - \gamma \br{F} + \alpha \br{FG}^* - \phi p_\mathrm{f} N \br{G}  \fgg^*,
	\end{equation}
where $\br{FG}^*,\fgg^*$ are functions of $\br{F}$ and $\phi$ (as shown in
\cref{fig:scatterplots_changes}a-c), and $\br{G}{=}1-\br{F}$. With these functions $\br{FG}^*$ and 
$\fgg^*$, the observed bistability is caused by a
classic double-well potential of the gradient system 
\cref{eq:dFdtclust2ss}. 
In this ODE, nonlinearities emerge due to the 
equilibrium dependence of the interface on forest area (affecting $\br{FG}^*$ 
and $\fgg^*$),
due to the segmentation of grass cells near and below the percolation threshold
(affecting  $\fgg^*$) and due to dependence of the ignition
rate on grass patch size (multiplying $\fgg^*$ with $\br{G}$).
In \exref{Fig.~S1}, 
we show the roots of \cref{eq:dFdtclust2ss}
using a nonparameteric fit of $\br{FG}^*(\br{F};\phi)$
and $\fgg^*(\br{F};\phi)$. These match well with
the steady states obtained via control (dot-dashed red).

If there is only one connected
component of grass cells, we have $\fgg{=}\br{FG}$, 
such that \cref{eq:dFdtclust2ss} simplifies to
	\begin{equation}\label{eq:dFdtclust1ss}
		\frac{\mathrm{d}\br{F}}{\mathrm{d}t} = \beta\br{G} - \gamma \br{F} + (\alpha - \phi p_\mathrm{f} N \br{G}) \br{FG}^*.
	\end{equation}
For homogeneous initial conditions (i.e. $\br{F}$ is about the same in different
large subsections of the domain), this approximation is expected to be valid for small $\br{F}$, where
most grass cells belong to the giant connected component.
\exref{Figure~S1} 
shows
the resulting steady states of \cref{eq:dFdtclust1ss} as a function of
$\phi$ and $\br{F}$ when only using the fit of $\br{FG}^*(\br{F};\phi)$ (dashed blue).
The approximation is good for landscapes with low forest cover ($\br{F}\lesssim0.2$). Above $\br{F}\approx0.2$, it fails because the
grassland breaks up into multiple clusters and fires are smaller than in case of a single 
cluster. \Cref{fig:scatterplots_changes}a--c already indicated that the single-cluster approximation is accurate for low forest cover since
$\fgg{\approx}\br{FG}$ for low $\br{F}$ in the scatterplots.

\subsection*{Resilience to perturbations} One can evaluate the right-hand side of \cref{eq:dFdtclust} for a landscape before and 
after application of a small perturbation, to determine if the perturbation
will be dampened or amplified under the dynamics. More precisely, 
we may define the sensitivity as
	\begin{align}\label{eq:jacob}
		\lambda_\mathrm{F}(X,\delta X):=\frac{\Delta\dot{\br{F}}(X,\delta X)}{\Delta\br{F}(X,\delta X)}=\frac{\dot{\br{F}}(X+\delta X)-\dot{\br{F}}(X)}{\br{F}(X+\delta X)-\br{F}(X)}, 
	\end{align}
for a landscape $X$ and a perturbation $\delta X$, where $\dot{\br{F}}$ is the
right-hand side of \cref{eq:dFdtclust}.
Negative values of $\lambda_\mathrm{F}$ correspond to 
dampening (negative feedback) and positive values to amplification (positive 
feedback). Given that the dynamics of \cref{eq:dFdtclust} are an approximate function of
$\br{F}$ only, the average of $\lambda_\mathrm{F}(X{+}\delta X)$ over naturally expected perturbations 
$\delta X$ (call this
$\bar{\lambda}_\mathrm{F}(X)$; see Methods, \cref{eq:meansens}) can be interpreted as
the approximate derivative 
$\mathrm{d}\dot{\br{F}}(\br{F})/\mathrm{d}\br{F}$, which
fully characterises the local stability of the landscape.
Therefore, the sign and magnitude of $\bar \lambda_\mathrm{F}(X)$ are indicators for the stability 
or criticality of a landscape. 
The dependence on only $\br{F}$ also implies that \cref{eq:dFdtclust} is a gradient system, such that 
$\bar{\lambda}_\mathrm{F}(X)$ is the concavity of the potential energy function 
at $X$, corresponding to the
classic potential-well metaphor of local resilience \cite{Scheffer2009,Scheffer2015}. 

\begin{figure}
	\includegraphics[width=\columnwidth]{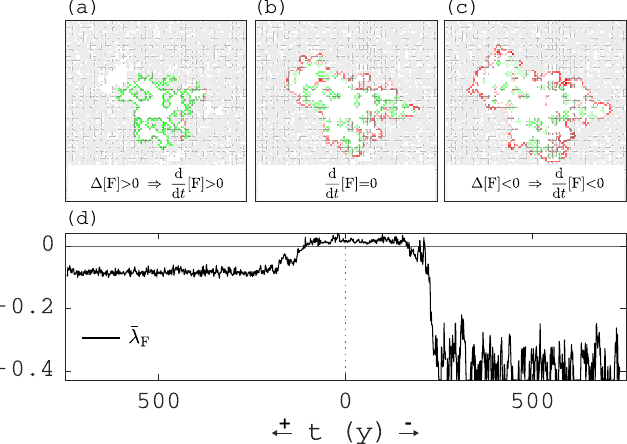}
	\caption{Resilience of forest-grass landscape to perturbations. (a--c) Spatially
		resolved contributions to forest gain (green) and loss (red) rates at the forest perimeter of the largest grass cluster: (b) for the saddle solution (where $\Delta_{\mathrm{F}}^{\mathrm{gain}}{\approx}\Delta_{\mathrm{F}}^{\mathrm{loss}}$), 
		(a) for a perturbation of the saddle with more forest at the perimeter (resulting in $\Delta_{\mathrm{F}}^{\mathrm{gain}}{>}\Delta_{\mathrm{F}}^{\mathrm{loss}}$), and (c) for a perturbation 
		of the saddle with less forest at the perimeter (resulting in $\Delta_{\mathrm{F}}^{\mathrm{loss}}{>}\Delta_{\mathrm{F}}^{\mathrm{gain}}$). (d) Sensitivity
		to perturbations (\cref{eq:jacob}) for all the traversed landscapes when tipping from the saddle: 
		down to the	grassland state ($t^-$), or, up to the forest state ($t^+$). The used landscapes
		correspond to the red vertical dashed line in \cref{fig:intro} and the rightmost columns in \cref{fig:gainlossterms,fig:scatterplots_changes}.}
	\label{fig:resilience}
\end{figure}

\Cref{fig:resilience}d shows $\bar \lambda_\mathrm{F}(X)$ 
for the traversed landscapes when tipping up to the forest or down to the grassland state (same landscapes as in \cref{fig:gainlossterms}b,d,f,g).
Comparison
of the magnitude of $\bar \lambda_\mathrm{F}(X)$ in the alternative stable states
reveals that (for parameters of \cref{fig:gainlossterms}d,f,g) the grassland state is more resilient than the forest state.
\Cref{fig:resilience}a--c shows the positive feedback for a
forest landscape with a hole of critical size (same landscape as shown at $t{=}0$ in \cref{fig:gainlossterms}b). Panel b shows which cells along the perimeter of the largest grass cluster contribute to the loss term $\Delta_\mathrm{F}^\mathrm{loss}$ (red) and the gain term $\Delta_\mathrm{F}^\mathrm{gain}$ (green). Panel a shows the effect of the perturbation obtained by converting the green cells to forest, which causes an increase in $\dot{\br{F}}$ (more green in panel a). Panel c shows the effect of the perturbation converting the red cells to grass, which causes a decrease in $\dot{\br{F}}$ (more red in panel c), illustrating the spatial distribution of the positive feedback.

One could, in principle, also test the effect of large perturbations, and whether
they will induce a transition to an alternative stable state, but 
it is not in general clear which perturbations are to be expected. 
However, in the special case of $\beta{=}\gamma{=}0$, when the
large perturbation is a single hole in contiguous forest, only the size of the hole matters,
and a simple expression for the critical size required to
tip abruptly to a non-forest state can be derived (see Methods, \cref{eq:crithole}). 

\subsection*{Comparison to mean-field approximations} 
\setcounter{paragraph}{0}
Our analysis of \cref{eq:dFdtclust} enabled us to obtain macroscopic steady states and dynamics without
making mean-field assumptions. 
\exref{Sections~S3} and \exref{S3}
derive a hierarchy of mean-field models for which we compare their predictions to our results to examine their validity. The simple mean field (\exref{Section~S3})
is unable to capture repeated fire extinction on a fast timescale and nearest-neighbour
spreading of forest and fire, leading to severe bias. 
When instead assuming timescale separation between forest and fire dynamics
and treating fire as a site percolation process in
landscapes with uniform random (i.e., spatially uncorrelated) placement of forest, we obtain the mean-field
approximation of \cref{eq:dFdtclust}:
	\begin{align}\label{eq:MF}
		\ddt\br{F} &=\beta\br{G}+4\alpha\br{F}\br{G}-\gamma\br{F}-\phi N p_\mathrm{f} \br{G}\fgg^\mathrm{u}.
	\end{align}
In \cref{eq:MF} we substituted the a-priori unknown forest perimeter $\br{FG}$ and the grassland-weighted perimeter $\fgg$ by their expressions assuming absence of correlations: $\br{FG}{\approx}4\br{F}\br{G}$ and $\fgg{\approx}\fgg^\mathrm{u}$ for
given $\br{F}$. The function $\fgg^\mathrm{u}(\br{F})$ is the quantity given in \cref{eq:defFGavcg} for 
uniformly randomly placed forest (see orange curve in \exref{Fig.~S3}a-c).
\cref{eq:MF} is bistable with stable high and low forest cover steady states over a wide range of parameters. It accurately
predicts the location of both stable steady states (blue line in 
\exref{Fig.~S2}).
Yet, its prediction of the threshold (unstable) steady state and the dynamics 
remains strongly biased. Indeed, away from the stable steady states,
the interplay of forest demography and fire-induced 
forest erosion creates landscapes in which forest is spatially aggregated, 
strongly violating the assumption of absence of correlations. 
This can be seen in 
\exref{Fig.~S3}a--c.
which shows that for given forest area $\br{F}$, 
the forest
perimeter of simulations, $\br{FG}$, lies below that predicted by the mean-field, 
$\br{FG}_\mathrm{mf}$, i.e.
	\begin{equation}
		\br{FG}<\br{FG}_{\mathrm{mf}}=4\br{F}(1-\br{F}),
	\end{equation}
implying that forest is
more spatially aggregated than assumed in the
mean field. Aggregation
results from forest spreading close to existing forest and 
from lower survival
of forest cells that are more exposed to fire (i.e., less aggregated).
Forest gain is smaller when aggregated (\exref{Fig.~S3}d--f)
due to the smaller perimeter. Aggregation reduces forest loss at low cover while it increases forest loss at high cover (\exref{Fig.~S3}d--f).
This is so because aggregation makes forest cells individually less exposed to fire, 
but collectively less effective at blocking fires, where the individual effect is 
dominant at low cover and the collective effect is dominant at cover values near and above the fire percolation threshold.

For our choice of parameters, the stable steady states and the lower
saddle-node bifurcation contain negligible spatial structure,
such that mean-field predictions are accurate. 
We derive their expressions below
for $\beta{\approx}0$ .

\paragraph{Stable steady states} By \cref{eq:MF}, the low-cover
steady state $\br{F}^*_{-}$ has to be approximately zero for $\beta{\approx}0$.
At high forest cover, loss due to fire is negligible, such that
the high-cover steady state $\br{F}^*_{+}$ can also be obtained from \cref{eq:MF}
(using $\beta{\approx}0$):
	\begin{equation}\label{eq:MFss}
		\br{F}^*_{-}=0,\qquad\br{F}^*_{+}=1-\frac{\gamma}{4\alpha}.
	\end{equation}
For the chosen parameters, we have $\br{F}^*_{+}{=}0.83$, which 
is in excellent agreement with simulations (\cref{fig:intro}).

\paragraph{Onset of bistability} At low forest cover, grass consists of a single cluster, such that
we can write in \cref{eq:MF} $\fgg^\mathrm{u}{=}\br{FG}{=}4\br{F}\br{G}$. Hence
(when $\beta{\approx}0$):
	\begin{equation}\label{eq:MF1cl}
		\ddt\br{F} =4\alpha\br{F}\br{G}-\gamma\br{F}-4\phi N p_\mathrm{f}\br{G}^2\br{F}.
	\end{equation}
An expression for the lower bifurcation point can be obtained by
finding the root of the derivative of the right-hand side of \cref{eq:MF1cl} with respect to $\br{F}$ at $\br{F}{=}\br{F}^*_{-}{=}0$, giving the relation $4\alpha-\gamma-4\phi Np_\mathrm{f}=0$. Rearranging this relation for $\phi N$, we can obtain the fire ignition rate above which 
tropical forests are bistable with grasslands:
	\begin{equation}\label{eq:minphiN}
		(\phi N)_{\mathrm{min}}=\frac{1}{p_\mathrm{f}}(\alpha-\frac{\gamma}{4}).
	\end{equation}
For the chosen parameters, 
$(\phi N)_{\mathrm{min}}{=}0.25$, which agrees well with simulations
(\cref{fig:intro}) and which corresponds to a maximum fire return
interval of $(\phi N)_{\mathrm{min}}^{-1}{=}4\mathrm{y}$.

\section*{Discussion}
\setcounter{paragraph}{0}
In this paper, we showed how nonlinear dynamics and bistability of tropical
forest emerge spontaneously from the patch-scale rules of forest and fire spread,
without assuming equations or thresholds for the effects of fire as in previous work.
Below, we first summarise our main results on structure and dynamics. Then we discuss
the importance of the emergent structure as indicated by comparison with mean-field 
approximations. Finally, we highlight the potential practical implications of our
results for resilience assessment and conservation.

\paragraph{Emergent structure and dynamics} Our simulations showed that spatial structure emerges due to
forest expansion and fire-induced damage at the forest perimeter.
As a consequence, the forest perimeter appeared in both the gain and loss side of our landscape-scale
balance equation of forest area, \cref{eq:dFdtclust}, where losses require weighting by adjacent grassland area.
Remarkably, when plotting the changes predicted by our balance equation versus forest area, using 
landscapes from the simulations, we found that they lie on an approximate curve 
(\cref{fig:scatterplots_changes}g--h). As this curve shows the change of forest area 
as a function of forest area, this means that the emergent macroscopic dynamics 
can be described by a simple ordinary differential equation, \cref{eq:dFdtclust2ss}.
In this emergent closed form of our balance equation, the perimeter quantities determine
the nonlinearities. Therefore, \cref{eq:dFdtclust2ss} elucidates how forest dynamics and bistability are linked
to the forest geometry that emerges from the patch-scale spreading rules.
Note that, as in previous work, \cref{eq:dFdtclust2ss} 
does not include fire
explicitly, because it does not contain equations for fire. This follows from timescale separation
between fire and vegetation dynamics, an assumption that
was already implicit in mean-field models  
(\citealp{Staver2012,Touboul2018,Li2019,Wuyts2019,Goel2020}; mean field in \citealp{Schertzer2014,Patterson2021}) and microscopic models (microscopic models in~\citealp{Schertzer2014,Patterson2021}) 
focusing on alternative stable states. 
However, previous work derived the implicit effect of fire
in closed form by relying on standard percolation theory, which assumes that occurrence of flammable patches is spatially uncorrelated \cite{Stauffer1994}. As we did not rely on percolation theory, 
but observed the closed form emerging in simulations (\cref{fig:scatterplots_changes}), we could avoid the biases
that affect previous work.

\paragraph{Evaluation of mean-field models}  We compared mean-field models against the emergent closed form of our balance equation 
to assess their validity (\exref{Fig.~S3}) and to show where spatial structure is important. 
This showed that mean-field models are in qualitative but not quantitative 
agreement with simulations: existence of bistability, but not its parameter range is robust to mean-field assumptions.
In particular, the simple mean field (\exref{Eq.~S7})
is strongly biased due to its failure to 
account for two phenomena that are present in the microscale model: 
(\emph{i}) spontaneous fire extinction on the fast timescale, leading to separated rapid fire spreading events, (\emph{ii}) nearest-neighbour spreading of fire and forest, 
leading to emergent aggregation of forest patches away from the steady states. The former
violates the mean-field assumption of large system size ($N{\to}\infty$) and the latter that of 
absence of correlations. That spatial structure affects steady states and dynamics
is well known \cite[e.g.][]{Keeling1999,Dieckmann2000}.
Even when addressing timescale separation and using results from percolation
theory for the effect of fire
(\exref{Eq.~S18} or \exref{Eq.~S22}), 
a large bias remains
except near the alternative stable states 
\exref{Fig.~S3}.
This is because standard percolation theory only considers lattice configurations with 
uniform random (i.e., spatially uncorrelated) placement of flammable sites while our forest-grass landscapes are shaped by past fires and vegetation dynamics.
As \cref{fig:gainlossterms}a,b (at $t{=}0$) shows, forest aggregation is particularly strong at the tipping
threshold for forest collapse, implying that mean-field models cannot be used to study abrupt forest
dieback. 
Despite their severe bias concerning forest dynamics and tipping, mean-field models are 
still useful for studying regimes with little structure, such as near the stable 
equilibria or for dynamics with uniform seed dispersal. This enabled us to derive expressions
for these equilibria (\cref{eq:MFss}) and the point of onset of bistability (\cref{eq:minphiN}). 
The latter result was not obtained by previous mean-field models because they did not 
include parameters that relate directly to fire 
\cite[][see \exref{Section~S7} 
for a suggested 
modification]{Schertzer2014,Staver2012} or they did not account for timescale separation in finite domains \cite{Hebert-Dufresne2018}.

\paragraph{Implications for resilience assessment and conservation} 
The link between geometry and dynamics implies that tropical forest resilience 
can be empirically estimated from its spatial structure. The spatial structure,
as captured by the perimeter quantities $\br{FG}$ and $\fgg$, can hence be treated as a measurable
parameter additional to the microscopic parameters. Microscopic parameters (given in \cref{tab:rates}) 
can be inferred from
remote-sensed data (as in \citealp{Hebert-Dufresne2018} or \citealp{Aleman2018}) or from experiments  
\cite[as for fire spread in][]{Cardoso2022}, while the perimeter quantities can be calculated
for any observed landscape. 
In regimes with negligible spatial structure, one can assess stability 
or resilience from the microscopic parameters alone, based on mean-field
results. E.g., if the onset point of bistability at low tree cover 
lies in the regime without spatial structure, as in our simulations,
one can directly estimate the minimum fire ignition rate for onset of bistability 
from the microscopic parameters (\cref{eq:minphiN}).
This expression then shows which
natural or abandoned degraded areas of low tree cover with fire ignition rate beyond this point will
not spontaneously recover to closed tropical forest. 
In regimes with spatial structure, the mean field is highly inaccurate (\exref{Fig.~S3}),
such that spatial structure needs to be considered in addition to the parameters. 
In particular, in our simulations, the tipping threshold obtains spatial structure
at higher fire ignition rates (\cref{fig:gainlossterms}a,b at $t{=}0$) and 
approaches the stable forest equilibrium
much more closely than in the mean field (\exref{Fig.~S2}). 
While this makes the mean field unsuitable for studying forest 
resilience and dieback, our balance equation (\cref{eq:dFdtclust}) 
does not have this limitation because it makes no assumption on spatial structure. 
We demonstrated how \cref{eq:dFdtclust} permits estimation of
the resilience of a landscape to perturbations, via $\lambda_\mathrm{F}$ (\cref{eq:jacob}).
In contrast to generic indicators
of resilience \cite{Scheffer2009,Boulton2022}, $\lambda_\mathrm{F}$ is
an indicator that can be obtained from a single landscape, and for which the contribution of each
relevant spatial process can be examined. Furthermore, landscape 
perturbations by human intervention can be evaluated, through sensitivity $\lambda_\mathrm{F}$, 
for how they will amplify or mitigate fire-vegetation feedback. 
Forest conservation/restoration may introduce targeted perturbations that 
most efficiently prevent resilience loss of high-cover states or induce resilience loss of low-cover 
states. For instance, in \cref{fig:resilience}a, forest dieback is averted by a perturbation that divides 
the largest grass 
cluster into smaller ones. It may thus be anticipated that maintenance or creation 
of barriers to fire spread will be essential here.

\vskip 4pt 
\noindent 
Future work could explore further realism, such as 
environmental heterogeneity, longer dispersal ranges, non-lattice geometry \cite[as in][]{Patterson2021}, 
inclusion of other tree types \cite[such as in savanna dynamics:][]{Staver2012,Schertzer2014,Touboul2018,VanLangevelde2003,Accatino2010,DOdorico2006,Baudena2010,Beckage2011,Hoyer-Leitzel2021,Patterson2021}, or vegetation-rainfall feedback \cite{Spracklen2018}. This
may result in additional relevant quantities in \cref{eq:dFdtclust}. 
Additionally, larger domain sizes may lead to more gradual transitions on the macroscopic scale \cite[][\exref{Section S6}]{Rietkerk2021}.

\section*{Methods}
\renewcommand{\theequation}{M\arabic{equation}} 
\renewcommand{\thefigure}{M\arabic{figure}}
\renewcommand{\thetable}{M\arabic{table}}
\setcounter{equation}{0}  
\setcounter{section}{0}  
\setcounter{figure}{0}  
\setcounter{paragraph}{0}

\paragraph{Details of the FGBA probabilistic cellular automaton}
The FGBA probabilistic cellular automaton is a minimal spatial 
stochastic process that models the joint dynamics of tropical vegetation 
and fire. It is an adapted
version of the BGT(A) model of ref. \citep{Hebert-Dufresne2018}. 
The modifications compared to ref. \citep{Hebert-Dufresne2018} are: 
(\textit{i}) it runs in continuous
time, (\textit{ii}) it includes a spontaneous forest mortality rate $\gamma$, (\textit{iii})  species $\mathrm{T}$ is labelled as $\mathrm{F}$,
consistent with other models of tropical vegetation dynamics \citep{Staver2012,Wuyts2019,Patterson2021}. Note that according to some definitions,
probabilistic continuous-time cellular automata are considered as interacting particle 
systems. In general, when studying
the stochastic dynamics of a number $n$ of interacting species on
a square lattice with $N$ cells, the state of the system can be represented
as 
\[
X:=(X_{1},X_{2},...,X_{N}),
\]
where $X_{i}$ is the label of the species that occupies cell $i$.
Each cell is occupied by exactly one of four possible species: grass, forest, burning and ash, with labels
$\mathrm{G}$, $\mathrm{F}$, $\mathrm{B}$ and $\mathrm{A}$. Transitions between states (species) occur in continuous time, resulting in a continuous-time Markov chain with a state space of size $n^{N}$. 
The reaction rules for transitions between states are shown in \cref{tab:rates},
where spontaneous conversions are shown on the left and conversions
due to nearest-neighbour interactions on the right (see also 
\cref{fig:FGBAplots}a). 

\begin{table}[h]
	\setlength{\tabcolsep}{2pt}
	\centering
	\caption{Reactions and rates (y$^{-1}$) in the CA.\label{tab:rates}}
	\begin{tabular}{|c|cl|cl|}
		\hline
		&\multicolumn{2}{c|}{Spontaneous}&\multicolumn{2}{c|}{Spread}\\
		\hline
		\parbox[t]{2mm}{\multirow{2}{*}{\rotatebox[origin=c]{90}{forest}}}
		&$\mathrm{G},\mathrm{A}\overset{\beta}{\to}\mathrm{F},$ & $\beta=2\cdot10^{-4}$ & 
		$\mathrm{GF},\mathrm{AF}\overset{\alpha}{\to}\mathrm{FF},$ & $\alpha_{\;}=3\cdot10^{-2}$\\
		&$\mathrm{F}\overset{\gamma}{\to}\mathrm{G},$&$\gamma=2\cdot10^{-2}$& &\\
		\hline
		\parbox[t]{2mm}{\multirow{3}{*}{\rotatebox[origin=c]{90}{fire}}}
		&$\mathrm{G}\overset{\phi}{\to}\mathrm{B},$ & $\phi=[0,2]\cdot10^{-4}$ & 
		$\mathrm{GB}\overset{\rho_\mathrm{g}}{\to}\mathrm{BB},$ & $\rho_\mathrm{g}=9\cdot10^6$\\ 
		&$\mathrm{B}\overset{\mu}{\to}\mathrm{A},$ & $\mu=10^6$ & $\mathrm{FB}\overset{\rho_\mathrm{f}}{\rightarrow}\mathrm{BB},$&$\rho_\mathrm{f}=1.11\cdot10^5$\\
		&$\mathrm{A}\overset{\lambda}{\to}\mathrm{G},$ & $\lambda=5$& &\\
		\hline
	\end{tabular}
\end{table}

\noindent The latter type of 
interaction occurs over each four nearest neighbour connections
of the indicated type. E.g., fire will spread into a given grass cell with
a rate $\rho_\mathrm{g}$ for each burning neighbour. For realistic timescales,
our parameters satisfy \cref{eq:fast-timescale1,eq:fast-timescale2},
which were empirically justified in \cite{Hebert-Dufresne2018}.
We borrow our notation from the moment closure
literature \cite[e.g.][]{Keeling1999,Dieckmann2000,Kiss2017,Wuyts2022}, writing
the global fraction of species with label $\mathrm{x}$ and the interface
between species with label $\mathrm{x}$ with label $\mathrm{y}$ respectively as
	\begin{equation}\label{eq:def:xxy}
		\br{x}:=\frac{1}{N}\sum_{i}^{N}\delta_\mathrm{x}(X_{i}),
		\quad
		\br{xy}:=\frac{1}{N}\sum_{i,j}^{N}\mathsf{A}_{ij}\delta_\mathrm{x}(X_{i})\delta_\mathrm{y}(X_{j}),
	\end{equation}
where both are normalised by $N$, $\delta$ is the Kronecker delta function ($\delta_\mathrm{x}(y){=}1$ if $y{=}\mathrm{x}$ and $0$ otherwise) and $\mathsf{A}\in\{0,1\}^{N\times N}$ 
the adjacency matrix. We simulated the cellular automaton via a Gillespie algorithm 
\citep{Gillespie2007} and used a domain
of $N{=}100{\times}100$ ($N{=}200{\times}200$ in \cref{fig:FGBAplots}) cells with periodic boundary conditions. 
%

\begin{figure}[t]
	\centering
	\includegraphics[width=1\columnwidth]{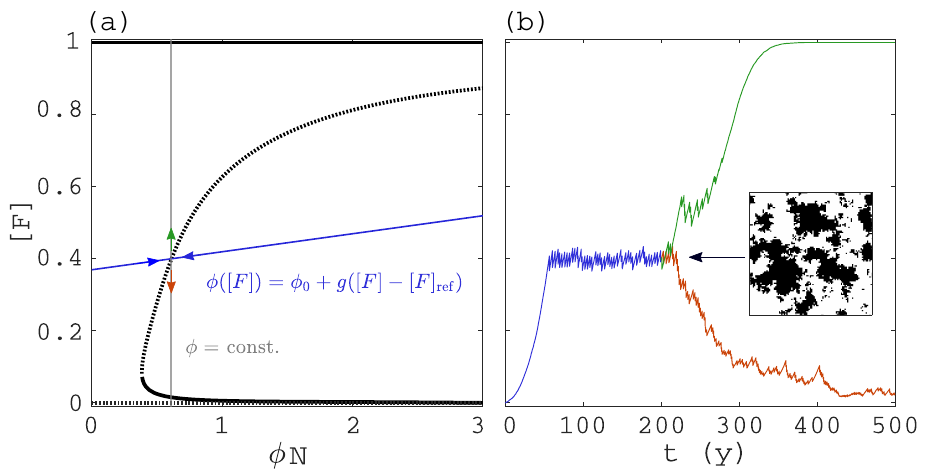}
	\caption{Feedback control applied to the CA without spontaneous mortality ($\gamma{=}0$): (a) the unstable steady state of the bifurcation diagram (dashed) was derived via feedback control by letting $\phi$ be a function of $\br{F}$ (blue line) such that it is stabilised, then obtaining $(\phi,\br{F})$	by averaging and repeating this for many $\br{F}_\mathrm{ref}$ (with appropriate $g$),	(b) a regular simulation with the same $\phi$ value (solid grey in (a)) and starting from the final state of the controlled simulation tips up or down depending on initial perturbations, (c) snapshot of the domain at the saddle for the control indicated in (a) (black: forest, white: grass). For other parameters, see \cref{tab:rates}.}
	\label{fig:ctrlFGBA}
\end{figure}

\paragraph{Noninvasive feedback control}
To study steady states regardless of their stability in a simulation, we apply noninvasive feedback control
\citep{SGNWK08,schilder2015experimental,barton2017control,neville2018shape}. 
To obtain the dependence of equilibria of $\br{F}$ on fire ignition rate $\phi$, we 
introduced an artificial stabilizing feedback loop of the form
	\begin{equation}
		\phi(t)=\phi_{0}+g(\br{F}(t)-\br{F}_{\mathrm{ref}}).\label{eq:control}
	\end{equation}
The factor $g$ is called the feedback control gain and is problem
specific. The property of noninvasiveness means that the controlled 
simulations have the same equilibria as
regular simulations \citep{BS13,SOW14,renson2017experimental}. This implies
that if one extracts the equilibrium values of the controlled simulation $(\phi^*,\br{F}^*)$,
one can use them to plot a 1-parameter bifurcation diagram of the simulation
without control.
\cref{fig:ctrlFGBA} shows
the control graphically. The 
feedback \cref{eq:control}, indicated in blue in \cref{fig:ctrlFGBA}a,
stabilises a steady state that is unstable in a regular simulation. This 
can be seen in \cref{fig:ctrlFGBA}b, where the unstable steady state is first
stabilised via control, after which the control is removed and a regular
simulation is started with the effective rate and the landscape (inset) 
obtained from the controlled
simulation. Depending on initial perturbations, the regular simulation gets
either attracted to the 100\% forest  state or to the low tree cover state.
When the controlled simulation is in equilibrium (steady part of the
blue curve in \cref{fig:ctrlFGBA}b), the steady state
values of $\br{F}$ are obtained via taking the time average, i.e.
	\begin{equation}\label{eq:effout}
		\overline{\br{F}}=\frac{1}{T}\int_{t_0}^{t_0+T}\br{F}(t)dt,
	\end{equation}
where $t_0$ is the time after which the dynamics have settled to a steady
state and $T$ the averaging time.
If $n_{\mathrm{G}\to\mathrm{B}}$ is the number of ignition events between $t=t_0$ and $t=t_0+T$, 
the steady state of $\phi$ is obtained by calculating the mean ignition
rate as $n_{\mathrm{G}\to\mathrm{B}}/T$ and dividing this by the mean number of grass
cells, such that
	\begin{equation}
		\bar{\phi}=\frac{n_{\mathrm{G}\to\mathrm{B}}/T}{\overline{\br{G}}},
	\end{equation}
where $\overline{\br{G}}$ is obtained as in \cref{eq:effout}.
When repeating this exercise for many $\br{F}_\mathrm{ref}$ values, 
one can get multiple points on the
unstable branch. Points on the stable branches can be obtained with
regular simulations. On the final selection of points, we applied
Gaussian process regression to obtain smooth curves and used 
moving block bootstrapping \citep{Kreiss2012} to obtain confidence
intervals.
One of the advantages of applying control is that one can obtain 
states for which one would have to wait prohibitively long in a 
regular simulation due to their instability. 

\paragraph{Forest loss due to a single fire} A fire in grassland cluster with index $j$ 
that reaches its interface with adjacent forest induces a forest loss that can be
approximated as follows. Consider a single forest cell $i$ 
located at the interface with grassland cluster $j$ with $\br{FG}_{i,j}$ number of 
neighbouring grass cells. When assuming that spreading events 
are independent, the probability that the forest cell gets burnt is
the complement of the probability that none of its neighbouring grass cells in cluster $j$
spread the fire to the forest cell:
	\begin{equation}
		q_{i,j}:=1-(1-p_\mathrm{f})^{[\mathrm{FG}]_{i,j}}\approx p_\mathrm{f} \br{FG}_{i,j},
	\end{equation}
where the approximation on the right is valid for small $p_\mathrm{f}$. Summing over all
forest cells at the interface of grassland cluster $j$, we obtain the expected
loss of forest per fire event as shown in \cref{eq:forest:loss}:
	\begin{equation}
		\Delta_{\mathrm{F},j}^{\mathrm{loss}}:=\sum_i q_{i,j}=p_\mathrm{f}\br{FG}_{j}.
	\end{equation}
This approximation also assumes that burning forest cells at the interface do not spread
the fire further, which also relies on $p_\mathrm{f}$ being small. For an evaluation of the
validity of this approximation in case of landscapes without spatial structure, see 
\exref{Fig.~S7}.

\paragraph{Critical hole size for an abrupt shift when $\gamma{=}\beta{=}0$}  When
there are no spontaneous transitions and we perturb a fully closed forest of
100\% cover by creating a hole with grassland, an expression can be obtained for the critical
hole size beyond which fire causes an abrupt shift to grassland. Using that grassland 
is a single cluster, \cref{eq:dFdtclust} becomes
	\begin{equation}\label{eq:dFdt:nospont}
		\frac{\mathrm{d}\br{F}}{\mathrm{d}t} = (\alpha - \phi N p_\mathrm{f} \br{G}) \br{FG},
	\end{equation}
which has two absorbing steady states $\br{F}^*_{\mathrm{l}}{=}0$ and $\br{F}^*_{\mathrm{h}}{=}1$, 
and an unstable
steady state at $\br{F}^*_{\mathrm{c}}{=}1-\frac{\alpha}{\phi N p_\mathrm{f}}$. The critical 
hole size is then the complement of the unstable steady state:
	\begin{equation}\label{eq:crithole}
		\br{G}_{\mathrm{c}}=\frac{\alpha}{\phi N p_\mathrm{f}},
	\end{equation}
which can also be written as 
$\br{G}_{\mathrm{c}}=\phi_1/\phi $,
where $\phi_1$ is the value of $\phi$
for which $\br{G}_{\mathrm{c}}=1$, or also
the lower limit of the bistablity range.

\paragraph{Sensitivity to perturbations} We estimated $\bar{\lambda}_\mathrm{F}$ in \cref{fig:resilience}d for
a given landscape by averaging \cref{eq:jacob} over realisations of different types of perturbations: 
	\begin{equation}\label{eq:meansens}
		\bar{\lambda}_\mathrm{F}= 
		\langle\frac{\sum_{i}w_i\Delta\dot{\br{F}}(X,(\delta X)_i)}{\sum_{i}w_i\Delta\br{F}(X,(\delta X)_i)}\rangle.
	\end{equation}
Each of $(\delta X)_i$ are perturbations that one might expect in simulations, such as
removal of a fraction of perimeter forest cells in the largest grass cluster or
spontaneous mortality/growth of forest. The weights $w_i$ are determined by the rates/probabilities of
occurrence of the perturbations. The $\langle\cdot\rangle$ denote an average over 64 realisations. 

\section*{Acknowledgements}
This work was supported by the UK EPSRC (grants EP/N023544/1 and EP/V04687X/1) and the EU Horizon 2020 TiPES project (grant 820970, output no. 171).

\bibliography{./lib}

\onecolumngrid
\renewcommand{\theequation}{S\arabic{equation}} 
\renewcommand{\thefigure}{S\arabic{figure}}
\renewcommand{\thesection}{S\arabic{section}}
\setcounter{equation}{0}  
\setcounter{section}{0}  
\setcounter{figure}{0}  

\begin{center}
	\fontsize{20}{24}\bfseries\scshape{Supplementary Materials}
\end{center}

\section{Supplementary figures referenced from the main text}
\noindent
\begin{figure}[H]
	\begin{minipage}{.45\columnwidth}
		\raggedleft
			\begin{center}
				\includegraphics[scale=1]{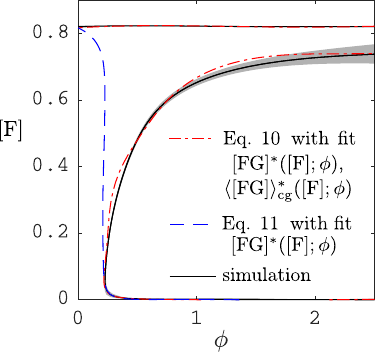}
			\end{center}
	\end{minipage}
	\hfill
	\begin{minipage}{.45\columnwidth}
		\raggedright
			\caption{Steady states of \cref{eq:dFdtclust2ss} (multicluster -- dot-dashed red) and of \cref{eq:dFdtclust1ss} (single cluster -- dashed blue) compared to controlled simulations (solid black with shading indicating two-standard deviation confidence interval of the mean). Parameters in the cellular automaton: $\gamma{=}0.02,\alpha{=}0.03,\beta{=}2\cdot10^{-4},\rho_\mathrm{g}{=}9\cdot 10^{6},\rho_\mathrm{f}{=}1.11\cdot10^{5},\mu{=}10^6,\lambda{=}5$.}\label{fig:fit_vs_ctrl}
	\end{minipage}
\end{figure}
\noindent
\begin{figure}[H]
	\begin{minipage}{.45\columnwidth}
		\includegraphics[scale=.8]{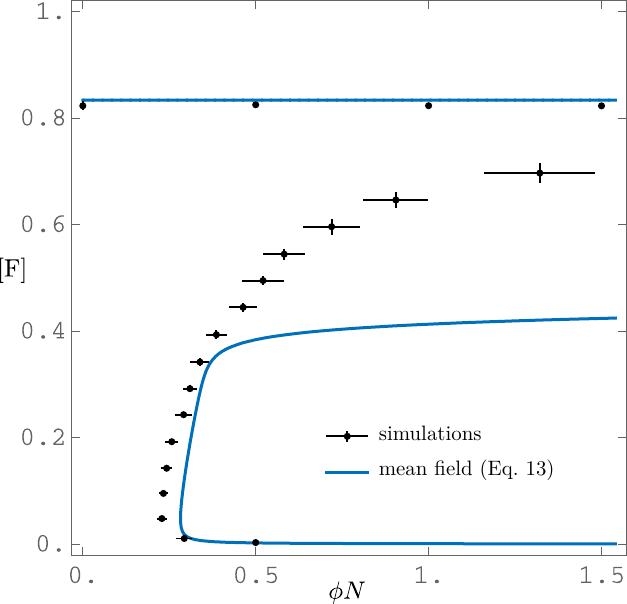}
	\end{minipage}
	\hfill
	\begin{minipage}{.45\columnwidth}
		\raggedright
			\begin{center}
				\caption{Comparison of steady state forest as a function of ignition rate in the time-separated mean-field and in simulations (dots with error bars: simulations, lines: mean field). A large difference between mean-field model and simulation occurs for the threshold steady state, while the mean-field model is accurate for high- and low-tree-cover alternative stable states. See \cref{sec:sup:MF2tp} and \cref{fig:compsim} for comparison of other scenarios and mean-field approximations.}\label{fig:compsim_simple}
			\end{center}
	\end{minipage}
\end{figure}
\noindent
\begin{figure}[H]
	\begin{center}
		\includegraphics[scale=.9]{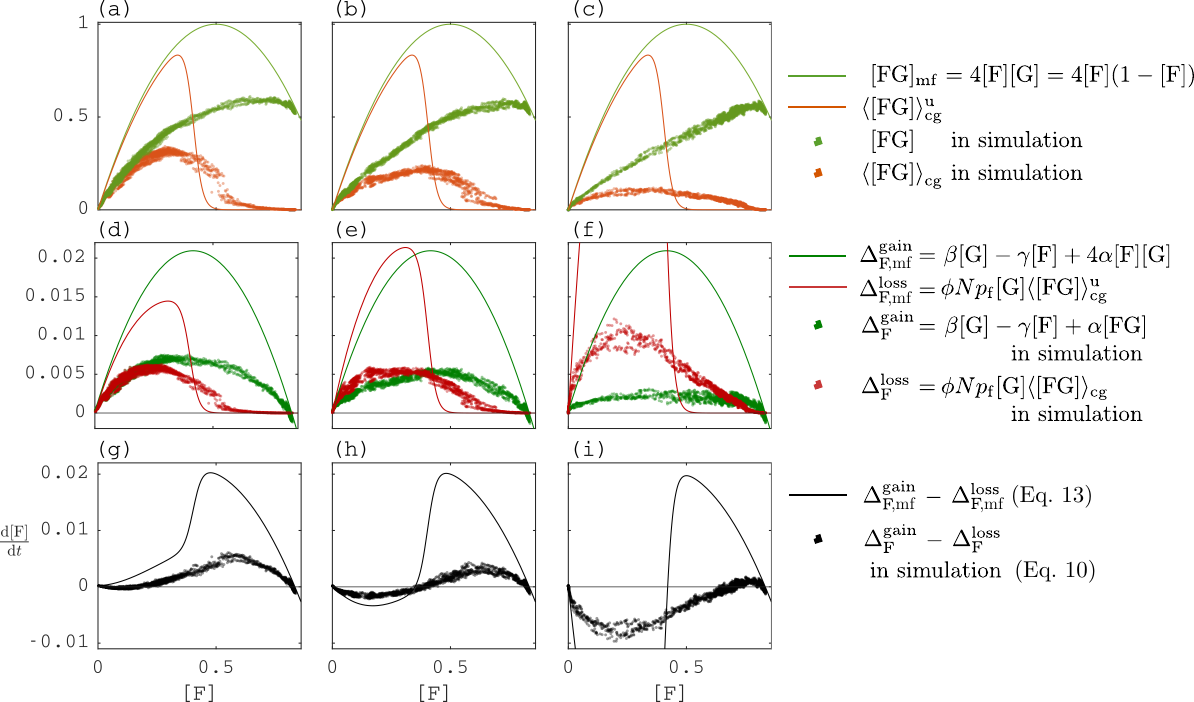}    
	\end{center}
		\caption{Emergent relations between key quantities and forest area $\br{F}$ compared to the mean field with site percolation (dots: simulations, lines: corresponding mean field quantities from \cref{sec:sup:MF2tp}): (a--c) forest perimeter $\br{FG}$ (green) and grassland-weighted forest perimeter $\fgg$ (orange), where green line equals $4\br{F}\br{G}$ and orange line was called $\fgg^u$ in the main article, (d--f) forest gain terms and loss terms in \cref{eq:slow:gain,eq:slow:loss}, (g--i) forest area rate of change $(\mathrm{d}/\mathrm{d}t)\br{F}$ from \cref{eq:dFdtclust}. Columns correspond to vertical dashed lines in \cref{fig:intro} ($\phi N{=}0.257,\phi N{=}0.38,\phi N{=}1.32$). Simulation results are identical to \cref{fig:scatterplots_changes}. Domain size: $N{=}100\mathrm{x}100$ cells. See \cref{sec:sup:MF2tp} for details of derivation for $\fgg^\mathrm{u}$ and $\br{FG}_\mathrm{mf}$.}\label{fig:sim3phivsMF_deco}
\end{figure}

\section{Relevant characteristics of the fire spreading process}\label{sec:GBA}

Before obtaining the mean-field equations for coupled vegetation and fire dynamics,
we analyse the fire spreading process in isolation. The insights from this section will enable us
to set up a mean-field model that constitutes the fairest comparison against the analysis
in the main text.

\subsection{Definition and mean field}\label{sec:MFGBA}  When we remove state $\mathrm{F}$ and its conversion rates to/from other types
($\alpha,\beta,\gamma,\rho_\mathrm{f}$) from the FGBA process, 
the dynamics show fire spread alone. We call this the \emph{GBA process}.
Writing $\mathrm{x}_i$ as shorthand for $\delta_\mathrm{x}(X_{i})$ (equalling $1$ if $X_i=\mathrm{x}$ and $0$ otherwise) and taking expectations in each cell $i$, 
we obtain equations for the rate of change of the expectation that cell $i$ is occupied by species $x\in\{\mathrm{G},\mathrm{B},\mathrm{A}\}$,

\noindent
	\begin{minipage}{.55\linewidth}
		\raggedright
		\begin{align}
			\ddt\bigE{\mathrm{G}_{i}} &= \lambda\bigE{\mathrm{A}_{i}}-\bigE{\mathrm{G}_{i}(\phi+\sum_{j\in{\cal N}(i)}\rho_\mathrm{g} \mathrm{B}_{j})},\nonumber &\\
			\ddt\bigE{\mathrm{B}_{i}} &= \phi\bigE{\mathrm{G}_{i}}-\mu\bigE{\mathrm{B}_{i}}+\bigE{\rho_\mathrm{g} \mathrm{G}_{i}\sum_{j\in{\cal N}(i)}\mathrm{B}_{j}},\nonumber &\quad\underset{\frac{1}{N}\sum_{i}^{N}}{\;\;\Longrightarrow}\\
			\ddt\bigE{\mathrm{A}_{i}} &= -\ddt\bigE{\mathrm{G}_{i}}-\ddt\bigE{\mathrm{B}_{i}}=\mu\bigE{\mathrm{B}_{i}}-\lambda\bigE{\mathrm{A}_{i}},\nonumber &
		\end{align}
	\end{minipage}%
	\begin{minipage}{.45\linewidth}
		\raggedleft
		\begin{align}\label{eq:MomeqsGBA2}
			\ddt\E{\br{G}} &= \lambda\E{\br{A}}-\phi\E{\br{G}}- \rho_\mathrm{g} \E{\br{GB}},\nonumber \\
			\ddt\E{\br{B}} &= \phi\E{\br{G}}-\mu\E{\br{B}}+\rho_\mathrm{g} \E{\br{GB}}, \\
			\ddt\E{\br{A}} &= \mu\E{\br{B}}-\lambda\E{\br{A}},\nonumber 
		\end{align}
	\end{minipage}
	\smallskip

\noindent 
where $\langle\cdot\rangle$ are ensemble averages, $\br{x}$ the domain fraction of species $\mathrm{x}$, 
and $\br{xy}$ the total number of 
neighbouring $\mathrm{xy}$ pairs divided by $N$, 
later referred to as the $\mathrm{xy}$ interface or $\mathrm{xy}$ perimeter. This set of
equations can be derived rigorously from the master equation \cite[e.g.][]{Tome2015}.
To go from individual (left) to population level (right), we summed 
over $i$ and divided by $N$, using \cref{eq:def:xxy}.
\Cref{eq:MomeqsGBA2} is not a
closed system. To close the system, we need to determine
all undetermined terms $\br{xy}$ on the right-hand side without creating
new unknowns. The simplest way to do this is to assume absence of 
pairwise correlations, i.e. $\E{\br{xy}}=4\E{\br{x}}\E{\br{y}}$. 
We take the additional assumption of $N\to\infty$, 
such that the law of large numbers applies and 
$\br{x}\to\E{\br{x}}$. These assumptions are valid when
all cells in an large domain interact with each other
at uniform contact rates of order $1/N$. This results in
the \emph{simple mean-field approximation} 
of the GBA process: 
	\begin{align}\label{MFeqsGBA0}
		\dot{\br{G}} &= \lambda\br{A}-\phi\br{G} -4\rho_\mathrm{g}\br{G}\br{B},\nonumber \\
		\dot{\br{B}} &= \phi\br{G}-\mu\br{B}+4\rho_\mathrm{g}\br{G}\br{B},\\
		\dot{\br{A}} &=\mu\br{B}-\lambda\br{A},\nonumber 
	\end{align}
where we also used the dot notation for time derivatives. Substituting
$\br{A}=1-\br{G}-\br{B}$ and taking only the independent equations, we finally obtain
	\begin{align}\label{eq:MFeqsGBA}
		\dot{\br{G}} &= \lambda(1-\br{G}-\br{B})-\phi\br{G} -4\rho_\mathrm{g}\br{G}\br{B}, \\
		\dot{\br{B}} &= \phi\br{G}-\mu\br{B}+4\rho_\mathrm{g}\br{G}\br{B}.\nonumber 
	\end{align}

We further focus on the case $\phi=0$, the reason for which will become clear below. When $\phi=0$, \cref{eq:MFeqsGBA} has two steady states, a trivial one at $(\br{G},\br{B})=(1,0)$
and one at $(\br{G},\br{B})=(\frac{\mu}{4\rho_\mathrm{g}},\frac{1-\mu/4 \rho _g}{1 +\mu/\lambda})$. The eigenvalues of the Jacobian of \cref{eq:MFeqsGBA} show that for $4\rho_\mathrm{g}/\mu>1$, the trivial steady states is a saddle and the 
non-trivial a spiral sink. For $4\rho_\mathrm{g}/\mu<1$, the trivial state state is a stable node and
the only physical solution. Hence, the steady states exchange stability at the transcritical
bifurcation at $4\rho_\mathrm{g}/\mu=1$. The GBA process with $\phi=0$ is equivalent to the SIRS spreading process in epidemiology \cite{Kiss2017}, which
represents spread of a disease in a population with waning immunity. 
Fire $\mathrm{B}$ plays the role of infected individuals.
Infections spread through a
population of susceptibles $\mathrm{G}$ at rate $\rho_\mathrm{g}$ per $\mathrm{GB}$ link. They subsequently acquire a state of immunity $\mathrm{A}$ at rate $\mu$, which can be lost at rate $\lambda$. The non-trivial steady state
corresponds to the endemic equilibrium and the transcritical bifurcation to the epidemic
threshold $R_0$.

\subsection{Extinction in finite systems} \label{sec:extinction}

\begin{figure}[h]
	\begin{center}
		\includegraphics[width=.9\columnwidth]{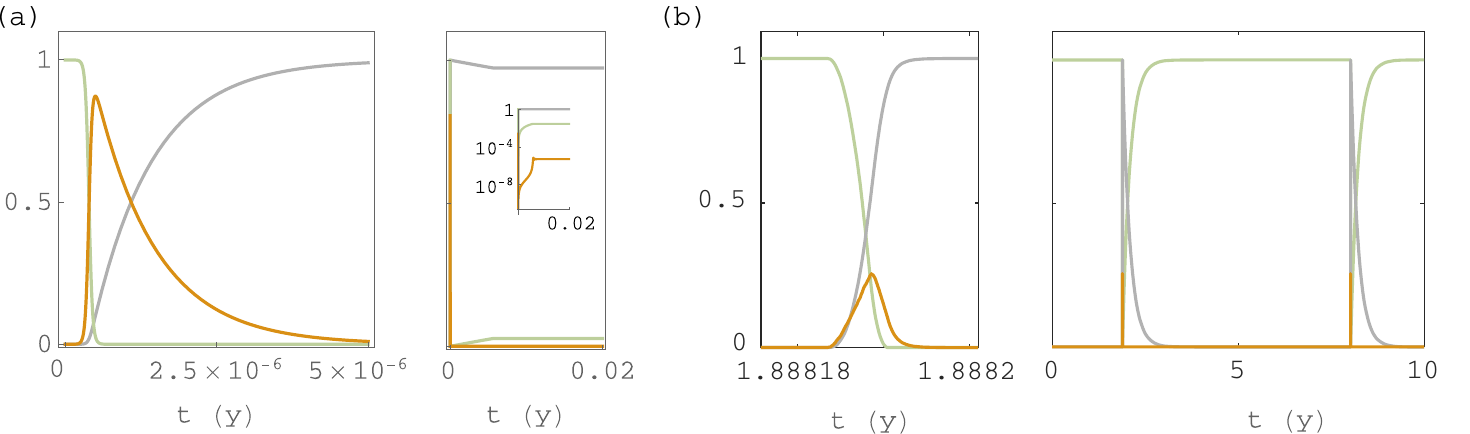}
		\caption{Time series of the GBA process ($\lambda,\phi\neq0$): (a) mean-field approximation, (b) simulation on a square lattice. G: green, B: orange, A: grey.}
		\label{fig:GBAt}
	\end{center}
\end{figure}

A well-known characteristic of this spreading process 
with $\phi=0$ and $\br{B}_0>0$ is 
that in finite systems, it goes extinct 
in finite time, even when $R_0>1$ \citep{Durrett1995}. This is so because stochastic excursions 
away from the non-trivial equilibrium will eventually reach the absorbing trivial state.
When the spontaneous
ignition rate $\phi>0$
and the time to extinction is much smaller than the typical waiting time
between ignition events, there are repeated fire events separated by extinction events.
The dynamics then effectively behave as a series of GBA processes with $\phi=0$ and $\br{B}_0>0$.
This is what we observe in cellular automaton simulations on a square lattice of
$100{\times}100$ cells for realistic parameters
(\cref{fig:GBAt}b, right panel). The mean-field approximation (\cref{eq:MFeqsGBA}), on the other hand, 
does not show extinction due to its assumption of $N\to\infty$. Instead, 
it shows a single pulse (\cref{fig:GBAt}a, left panel)
after which a high-ash low-grass and non-zero fire steady state (the endemic equilibrium) is reached 
(\cref{fig:GBAt}a, right panel).
In the case $\phi=0$, the required lattice size to avoid extinction with high probability depends on the initial
conditions \cite{Durrett1995}, but for realistic parameter ranges, it is unrealistically large. 
This can be understood as follows.
\begin{itemize}
	\item When the initial condition is a single fire, at least one cell has to keep on
	burning until the density of grass has regrown to a level sufficient for a new wave to
	propagate. This translates into the condition $(L/\Delta x)^2 \exp(-\mu/\lambda)\ge{\cal O}(1)$, such that
	$L\ge{\cal O}(10^{4\cdot10^4})$ for our parameters (taking a grid size of $\Delta x=0.03\text{km}$ 
	as in \cite{Hebert-Dufresne2018}).
	\item  When initial conditions are such that
	a band of the domain is immune at the start, a single fire can keep on burning
	by crossing the domain repeatedly \cite{Durrett1995}. 
	When assuming $\rho_\mathrm{g}\gg\mu$ and using that waiting times
	between spreading events are exponentially distributed with mean 
	$1/\rho_\mathrm{g}$, a fire will spread throughout the domain
	in a time of the order $\tau\approx L/(\rho_\mathrm{g}\Delta x)$. 
	For there to be sufficient regrowth of grass on this time scale,
	we need $L/(\rho_\mathrm{g}\Delta x)\approx1/\lambda$, or $L\approx\rho_\mathrm{g}\Delta x/\lambda$. 
	For the parameters we have chosen, 
	this means $L={\cal O}(10^4)$km, i.e.
	the order of magnitude of the earth's circumference,
	which is drastically smaller than the above estimate yet still
	impractically large. 
\end{itemize}
Taking more conservative estimates
for fire spreading rates or taking account of
a small positive fire ignition rate $\phi={\cal O}(\lambda/N)$ for the
initial condition with a single burning cell,
this may be decreased by an order
of magnitude, i.e. the size of a continent or country. 
Still, 
in reality, extinction will occur on smaller scales
due to spatiotemporal heterogeneity of forcing parameters as a
consequence of climatic seasonality or existence of natural or artificial
boundaries (such as forests), leading to a lower effective system size.
Hence, in any real system, repeated extinction 
and system-scale oscillations are to be expected.

\subsection{Percolation analysis of a single fire event}\label{sec:percsingle} 
\begin{figure}[h]
	\centering
	\includegraphics[width=1\columnwidth]{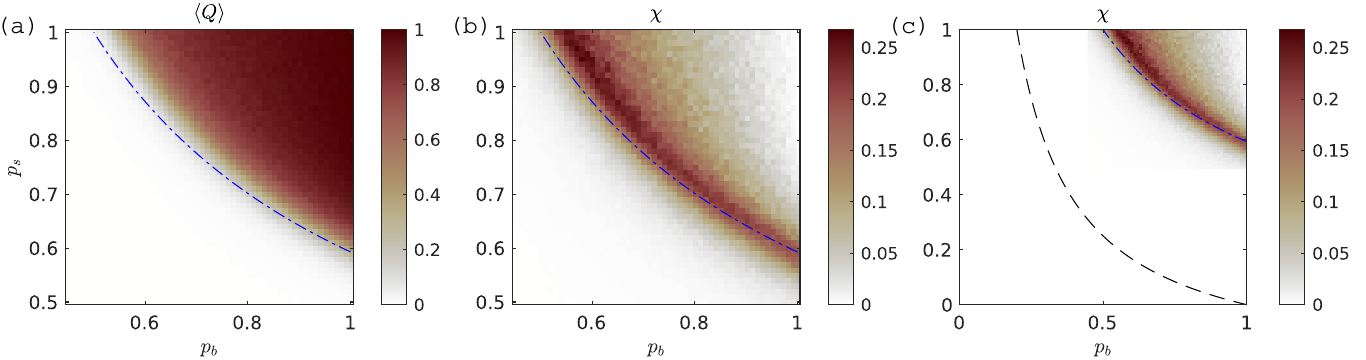}
	\caption{GBA process with $\phi{=}\lambda{=}0$ (equivalent to square lattice SIR spreading): $\E{Q},\chi$ versus bond occupation probability $p_b:=p_\mathrm{g}$ (\cref{eq:occupation:prob}) and site occupation probability $p_s{=}\br{G}_0$.  (a) Mean cumulative probability of being burnt $\E{Q}$ (expectation of \cref{eq:Qdef}). (b) Susceptibility $\chi$ (\cref{eq:chidef}). (c) Susceptibility $\chi$ compared to the mean-field percolation threshold  $p_{s,\mathrm{mf}}$, given in \cref{eq:meanGBAthresh} (dashed black). The dash-dotted blue line indicates the location of the infinite-size percolation threshold for uncorrelated mixed site-bond percolation (taken from \citep{Tarasevich1999}). The GBA model's percolation threshold lies at higher values (b) due to spatial correlation of $p_\mathrm{g}$ as explained in the text. The colour scale was taken from \cite{Crameri2018}. See \cref{fig:percolation} for more detail.}\label{fig:percolationshort}
\end{figure}

Therefore, a single fire 
in realistically sized systems corresponds to the case  
$\phi=0$, starting with a single burning cell. Using that the 
regrowth of grass occurs on a much slower time scale, we can further
also set $\lambda=0$ in our following analysis.
The GBA process with $\phi,\lambda=0$ is equivalent to susceptible-infected-recovered
(SIR) epidemic spreading \citep{Kiss2017}.
The final size of the epidemic in SIR epidemic spreading
on a lattice shows a continuous phase transition (CPT) at a critical spreading rate
$\rho_\mathrm{g}$ and scaling laws near the critical point obey those of
the ordinary percolation universality class \citep{Tome2010}. 
\Cref{fig:percolationshort} shows mean quantities for SIR epidemic spreading
on a square lattice in a range of infection probabilities and initial
number of immune individuals, which are spatially uniformly distributed.
In particular, we show that
SIR epidemic spreading is a type of \emph{mixed site-bond percolation}, with
bond occupation probability given by $p_{b}:=p_\mathrm{g}=\rho_\mathrm{g}/(\rho_\mathrm{g}+\mu)$
(which is fixed at $0.9$ in the main text, as in ref. \citep{Hebert-Dufresne2018})
and site occupation probability given by $p_{s}:=\br{G}_0$, i.e. the
initial fraction of cells that are grass in fire spreading, or the
complement of the initial fraction of immune individuals in epidemic
spreading (with the rest being susceptible).  In
\cref{fig:percolationshort}, we record the \emph{mean cumulative probability
	of being burnt} $\E{Q}$ and the \emph{susceptibility} $\chi$ of
$\E{Q}$, defined as

\noindent
	\begin{minipage}{.5\textwidth}
		\begin{align}\label{eq:Qdef}
			Q&:=\frac{\br{A}^*-\br{A}_0}{\br{G}_0},
		\end{align}
	\end{minipage}
	\begin{minipage}{.5\textwidth}
		\begin{align}\label{eq:chidef}
			\chi&=\frac{\E{Q^2}}{\E{Q}},
		\end{align}
	\end{minipage}

\noindent where $\br{\cdot}_0$ denotes initial value and $\br{\cdot}^*$ final value. 
For $N{\to}\infty$, $\E{Q}$ converges to the \emph{percolation probability} $P_\infty$, which is
the probability that a grass cell belongs to the giant connected component. 
We use the location where $\chi$ peaks as 
an estimate of the percolation threshold \cite{Hebert-Dufresne2019,Stauffer1994}. 
When $p_s=\br{G}_0=1$, we have pure bond percolation
and when $\rho_\mathrm{g}/\mu\to\infty$, we have pure site percolation.
The percolation threshold for standard mixed site-bond percolation
(from \citep{Tarasevich1999}) is shown in \cref{fig:percolationshort} 
with a dot-dashed blue curve.
The pure bond percolation threshold (when $\br{G}_0=1$)
of SIR epidemic spreading occurs at higher $p_b$ than in standard bond
percolation ($p_{b}\approx0.538>0.5$) because the possibility of spreading
to multiple neighbours makes the bond occupation probability spatially autocorrelated,
as shown by \citep{Tome2010}. The pure site percolation limit shows the classical
value (for the square lattice) of $\br{G}_0\approx0.593$.
From \cref{eq:MFeqsGBA} 
(with $\phi=\lambda=0$), we
can obtain the mean-field percolation threshold $p_{s,\mathrm{mf}}$ by finding where
the trivial state becomes unstable in \cref{eq:MFeqsGBA}, which is given by
	\begin{equation}
		4\br{G}_0\frac{\rho_\mathrm{g}}{\mu}=1\;\implies\; p_{s,\mathrm{mf}}=\br{G}_0=\frac{1}{4}\left(\frac{1}{p_b}-1\right).\label{eq:meanGBAthresh}
	\end{equation}
As shown in \cref{fig:percolationshort}c, the mean-field approximation
shows a large bias towards lower values.

\paragraph{Implications for the FGBA process} On landscapes with forest, fires 
can be blocked (albeit imperfectly) by forest cells. These landscapes obtain a
steady state shape due to the shaping processes of forest demography and fire. Hence, 
results from the spatially uniform $\br{G}_0$ above do not apply to percolation
effects in the full FGBA process. That is, when fire spreads on landscapes
with forest, the critical point for pure site percolation ($\rho_\mathrm{g}/\mu\to\infty$, or 
$p_\mathrm{g}\to1$) will in general depend not only on the site occupation probability
$\br{G}_0$ but also on the spatial correlation function of site occupation.
For the idealised case of fire-proof forest ($p_\mathrm{f}=0$), fire percolation on real landscapes is then equivalent to \emph{correlated
	mixed percolation}, where correlations in bond occupation probability occur due
to the spreading process, and correlations in site occupation probability occur
due to the nonrandom spatial structure of the landscape. When $p_\mathrm{f}>0$, the 
spreading process becomes a \emph{heterogeneous (correlated) bond percolation processes}, i.e.
a percolation process in which
fire spread on grass occurs with bond occupation probability $p_\mathrm{g}$ and on forest
with bond occupation probability $p_\mathrm{f}$. The possibility of spreading on forest
decreases the percolation thresholds compared to the correlated mixed percolation
limit of $p_\mathrm{f}\to0$. This decrease
is expected to be small because forests do not spread fires well ($p_\mathrm{f}\approx0$).

\section{Simple mean field of joint forest and fire spread}\label{sec:sup:MF}
\begin{figure*}[h]
	\centering
	\includegraphics[width=1\columnwidth]{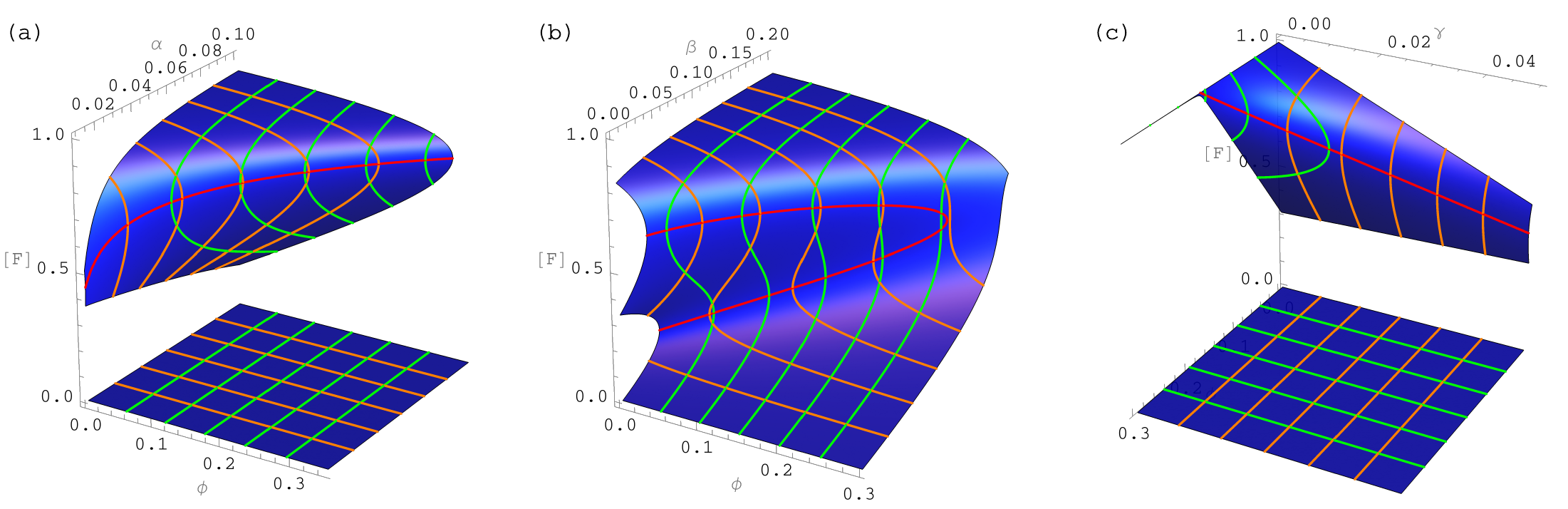}
	\caption{Steady states and bifurcations of forest cover $\br{F}$ in the simple mean field of the FGBA process (blue: steady state manifold, green/orange contours at fixed axis values, red: saddle-node bifurcations): (a) versus fire ignition rate $\phi$ and forest spreading rate $\alpha$, (b) versus fire ignition rate $\phi$ and spontaneous forest growth rate $\beta$, (c) versus fire ignition rate $\phi$ and spontaneous forest mortality rate $\gamma$. Due to its large bias, the simple mean field shows different bistability ranges than the simulations. We set $p_\mathrm{g}{=}0.25$ (requiring a $\rho_\mathrm{g}$ that is 27x smaller than in simulations)
		such that bistability ranges are visible (remaining parameters are as in \cref{tab:rates}).}
	\label{fig:FGBA_simpleMF}
\end{figure*}

When we follow the same steps as in \ref{sec:MFGBA}, we obtain the \emph{simple mean-field approximation} of the FGBA process:
	\begin{align}
		\dot{\br{G}} &= \lambda\br{A}-\phi\br{G} -4\rho_\mathrm{g}\br{G}\br{B}-\beta\br{G}-4\alpha\br{F}\br{G}+\gamma\br{F},\nonumber \\
		\dot{\br{F}} &= \beta\br{A}+4\alpha\br{F}\br{A}-4\rho_\mathrm{f}\br{F}\br{B}+\beta\br{G}+4\alpha\br{F}\br{G}-\gamma\br{F},\label{eq:MFeqs}\\
		\dot{\br{B}} &= \phi\br{G}-\mu\br{B}+4\rho_\mathrm{f}\br{F}\br{B}+4\rho_\mathrm{g}\br{G}\br{B},\nonumber \\
		\dot{\br{A}} &=\mu\br{B}-\lambda\br{A}-\beta\br{A}-4\alpha\br{F}\br{A}.\nonumber 
	\end{align}
The simple mean field shows bistability of tree cover (first shown by ref. \citep{Hebert-Dufresne2018} for $\gamma=0$) in ranges of all
parameters that are expected to show considerable spatial 
heterogeneity in a given ecosystem: $\alpha,\beta,\gamma,\phi$ (\cref{fig:FGBA_simpleMF}). 
However, despite being qualitatively correct, it shows a large bias compared to simulations. 
For the
parameter ranges of our simulations, it has no non-trivial solution for
positive fire ignition rate $\phi$, so we had to choose different parameter values
to find its bistability range. This bias is due to the inability of the simple mean
to capture two effects: repeated fire extinction on a fast timescale and the spatial nature of the two
spreading processes.
In the following section, we derive alternative mean-field models that partially correct 
for these biases.

\section{Two-timescale mean field of joint forest and fire spread} \label{sec:sup:MF2t}
Here,
we derive an alternative mean-field model that 
takes account of separation of fire events in systems with realistic sizes, assuming 
that fire spread occurs on a much faster time scale than forest spread.
This means that we can consider the fire spreading process in isolation
with $\phi=\lambda=0$ (as argued in \cref{sec:GBA}), and take the asymptotic amount 
of forest burnt by a single fire before extinction on the fast time scale as forest mortality per fire event on the slow time scale.

\subsection{Well-mixed fire and forest} \label{sec:sup:MF2twm}

We start with the simplest case, where both vegetation and fire mix uniformly, which is one way to conform with the mean-field assumption
of absence of correlations. 

\subsubsection{Fast process: forest loss due to a single fire} 
\begin{figure*}[h]
	\centering
	\includegraphics[width=.8\columnwidth]{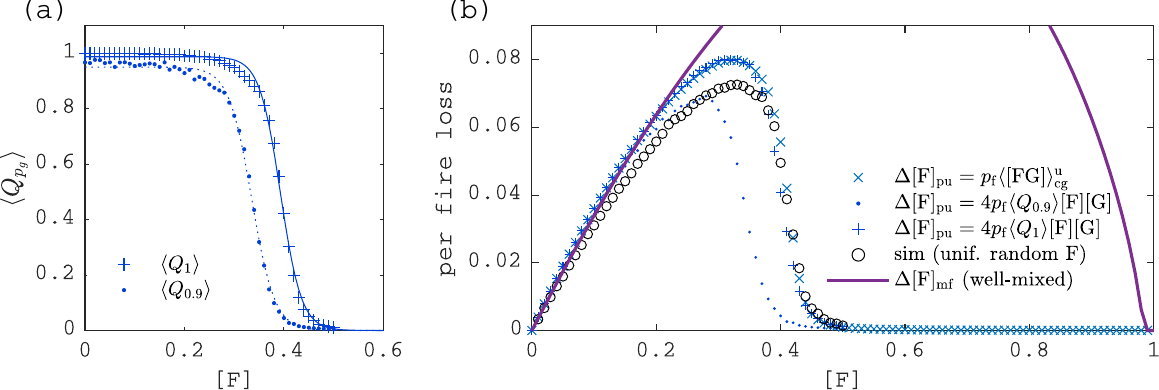}
	\caption{Grass burning probabilities and forest loss per fire in landscapes without spatial structure: (a) probability that a grass cell burns $\E{Q_{p_\mathrm{g}}}$, for $p_\mathrm{g}=1$ (`$+$') and for $p_\mathrm{g}=0.9$ (`$\cdot$'), together with fits to logistic functions; (b) loss per fire estimated from grassland-weighted forest (`$\times$', \cref{eq:losspfpu1}), from $\E{Q_{0.9}}$ (`$\times$', \cref{eq:losspfpu2}), from $\E{Q_{1}}$ (`$+$', \cref{eq:losspfpu2}), by assuming uniform mixing (purple curve, \cref{eq:dFtsepMFmixedall}), and measured in fire simulations with uniform random placement of forest (`$\circ$').}\label{fig:lossperfire}
\end{figure*}

On the fast time scale, we can set all small parameters related to forest demography, grass regrowth and fire ignition to zero 
($\alpha{=}\beta{=}\gamma{=}\lambda{=}\phi{=}0$), such that we obtain
	\begin{align}\label{eq:MFfgba_fast}
		\ddt\br{G} &= -4\rho_\mathrm{g}\br{G}\br{B},\nonumber \\
		\ddt\br{F} &= -4\rho_\mathrm{f}\br{F}\br{B},\\
		\ddt\br{B} &= -\mu\br{B}+4\rho_\mathrm{f}\br{F}\br{B}+4\rho_\mathrm{g}\br{G}\br{B},\nonumber \\
		\ddt\br{A} &= \mu\br{B},\nonumber 
	\end{align}
where the products arise from the well-mixedness assumption as before.
By rewriting the equations for $\ddt\br{G},\ddt\br{F},\ddt\br{A}$ as
	\begin{align}
		-\frac{1}{4\rho_\mathrm{g} \br{G}}\ddt \br{G} &= -\frac{1}{4\rho_\mathrm{f} \br{F}}\ddt \br{F} = \frac{1}{\mu}\ddt \br{A} = \br{B},
	\end{align}
we can obtain $\br{G}$ and $\br{F}$ as a function of $\br{A}$ via separation of variables and integration:
	\begin{align}\label{eq:GandFvsA}
		\br{G}(t)=\br{G}_0\exp\bigl(-\frac{4\rho_\mathrm{g}}{\mu}\br{A}(t)\bigr), \quad& \br{F}(t)=\br{F}_0\exp\bigl(-\frac{4\rho_\mathrm{f}}{\mu}\br{A}(t)\bigr).
	\end{align}
Substituting \cref{eq:GandFvsA} into the equation for $\ddt\br{A}$ in \cref{eq:MFfgba_fast}
and setting the time derivative to zero, we obtain an implicit relation of the asymptotic 
amount of vegetation burnt:
	\begin{align}\label{eq:asymptoticA}
		\br{A}^*&=1-\br{G}^*-\br{F}^*=1-\br{G}_0\exp\bigl(-\frac{4\rho_\mathrm{g}}{\mu}\br{A}^*\bigr)-\br{F}_0\exp\bigl(-\frac{4\rho_\mathrm{f}}{\mu}\br{A}^*\bigr).
	\end{align}
When taking an initial state consisting of only grass and forest, that is $\br{G}_0=1-\br{F}_0$,
$\br{A}^*$ can be found numerically as a function of $\rho_\mathrm{g}/\mu,\rho_\mathrm{f}/\mu$ and $\br{F}_0$.
This can in turn be used to obtain the total amount of forest lost due to a single fire,
from \cref{eq:GandFvsA},
	\begin{align}\label{eq:dFtsepMFmixedall}
		\Delta \br{F}_{\mathrm{mf}}:=\br{F}_0-\br{F}^*=\br{F}_0\Bigl(1-\exp\bigl(-\frac{4\rho_\mathrm{g}}{\mu}\br{A}^*\bigr)\Bigr),
	\end{align}
where subscript $\mathrm{mf}$ denotes mean field. A plot of \Cref{eq:dFtsepMFmixedall}
versus $\br{F}$ is given in \cref{fig:lossperfire} (solid purple curve), which shows
that the percolation threshold -- where forest starts blocking fire -- lies at
unrealistically low grass cover, as was also the case for perfectly blocking forest
(\cref{fig:percolationshort}c, dashed line).

\subsubsection{Slow processes: forest demography and fire damage} Now, we can define the mean-field forest gain and loss terms on the slow time scale as
	\begin{align}
		\Delta_{\mathrm{F},\mathrm{mf}}^{\mathrm{gain}} &:= \beta\br{G}+4\alpha\br{F}\br{G}-\gamma\br{F},\label{eq:MFtsepgain}\\
		\Delta_{\mathrm{F},\mathrm{mf}}^{\mathrm{loss}} &:= \phi N \br{G}\Delta \br{F}_{\mathrm{mf}},
	\end{align}
such that the final mean-field model becomes
	\begin{align}
		\ddt\br{F} &= \Delta_{\mathrm{F},\mathrm{mf}}^{\mathrm{gain}}-\Delta_{\mathrm{F},\mathrm{mf}}^{\mathrm{loss}}\nonumber\\
		&= \beta\br{G}+4\alpha\br{F}\br{G}-\gamma\br{F}-\phi N \br{G}\Delta \br{F}_{\mathrm{mf}},\nonumber\\
		&= \beta(1-\br{F})+4\alpha\br{F}(1-\br{F})-\gamma\br{F}-\phi N (1-\br{F})\br{F}\Bigl(1-\exp\bigl(-\frac{4\rho_\mathrm{g}}{\mu}\br{A}^*(\br{F})\bigr)\Bigr).
	\end{align}
The steady states are shown in
\cref{fig:compsim} in solid purple for the same parameters
as those used in the simulations of the main text.
For low and high tree cover, it
reproduces the steady states fairly accurately, but
unlike the simulations, it shows no wide saddle in between.
Hence, while this is an improvement compared to the simple mean field,
there is still a large bias at intermediate tree cover. To address this bias,
we need drop the assumption of uniform mixing for fire spread.

\subsection{Spatial fire percolation and uniformly randomly placed forest}\label{sec:sup:MF2tp} While the uniform mixing
assumption may be
ecologically justified for forest spread in case of species with 
long-range seed dispersal, it is much harder to justify for fire spread, 
which is fundamentally a local contagion process. Therefore, we
aim to take into account the effects of fire as a percolation process
while still assuming absence of spatial correlations between forest cells.
Because the percolation process affects forest loss, this only affects the loss function. 
Earlier mean-field models 
\cite{Staver2012,Schertzer2014,Patterson2021} accounted for the effects of
fire percolation by making fire-affected rates threshold
functions of tree cover, such as those shown in \cref{fig:lossperfire}a, while assuming that vegetation remains spatially uncorrelated  \cite[for derivation, see][]{Schertzer2014}. Therefore the mean-field analyses presented here provide the fairest points of comparison. 

To estimate the loss due to fire, we will show two alternative 
approaches. The first is equivalent to our approach in the main text, using
grassland-weighted forest perimeter to estimate exposed forest.
The second estimates exposed forest via standard results from percolation theory, which are valid here due to the assumption of uniform random placement of forest. 

\begin{figure*}[h]
	\centering
	\includegraphics[width=1\columnwidth]{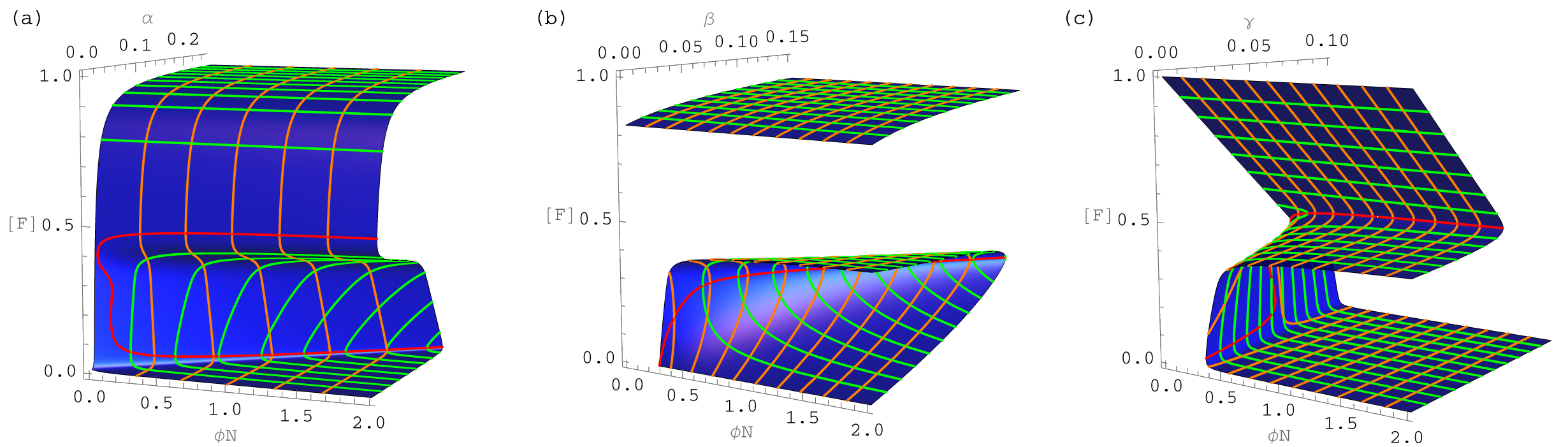}
	\caption{Steady states and bifurcations of forest cover $\br{F}$ in the two-timescale mean field of the FGBA process via \cref{eq:MF2t_FGcg} (blue: steady state manifold, green/orange contours at fixed axis values, red: saddle-node bifurcations) as a function of (scaled) fire ignition rate $\phi N$ and: (a) forest spreading rate $\alpha$, (b) spontaneous forest growth rate $\beta$, (c) spontaneous forest mortality rate $\gamma$. Parameters others than the ones on the axes are the same as those chosen in simulations -- see \cref{tab:rates}.}
	\label{fig:FGBA_MFFGcg}
\end{figure*}

\begin{figure*}[h]
		\includegraphics[width=.8\columnwidth]{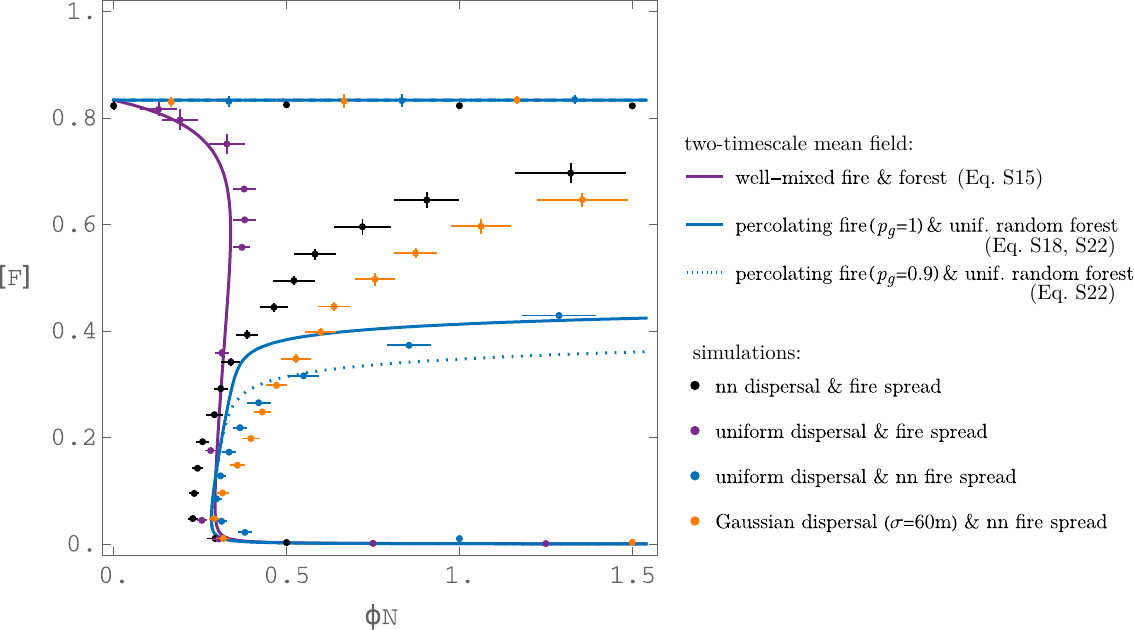}
		\caption{Comparison of steady state forest as a function of ignition rate in the time-separated mean-field and in simulations (dots with error bars: simulations, lines: mean field). See legend for details. Note, `uniform random' refers to the placement of forest cells in the domain.}\label{fig:compsim}
\end{figure*}

\begin{enumerate}
	\item Using the\emph{ grassland-weighted forest perimeter} $\fgg$ (see \cref{eq:defFGavcg}). According to this approach, fires spread perfectly to the forest perimeter, where a fraction of the forest is burnt. The difference
	with the main text is that the landscapes
	in which fire spreads have uniform random placement of forest. We indicate this difference below
	by the superscript $\mathrm{u}$ in $\fgg^\mathrm{u}$. The resulting
	loss per fire is
		\begin{equation}\label{eq:losspfpu1}
			\Delta \br{F}_\mathrm{pu}= p_\mathrm{f} \fgg^\mathrm{u},
		\end{equation}	
	such that the loss function is
		\begin{equation}\label{eq:losspu1}
			\Delta_{\mathrm{F},\mathrm{pu}}^{\mathrm{loss}}=\phi N \br{G}\Delta \br{F}_\mathrm{pu}=\phi N p_\mathrm{f} \br{G}\fgg^\mathrm{u},
		\end{equation}	
	where subscript $\mathrm{pu}$ refers to percolation on a square lattice with uniform random
	placement of forest. The final mean-field model is 
		\begin{align}\label{eq:MF2t_FGcg}
			\ddt\br{F} &= \Delta_{\mathrm{F},\mathrm{mf}}^{\mathrm{gain}}-\Delta_{\mathrm{F},\mathrm{pu}}^{\mathrm{loss}},\nonumber\\
			&=\beta(1-\br{F})+4\alpha\br{F}(1-\br{F})-\gamma\br{F}-\phi N p_\mathrm{f} (1-\br{F})\fgg^\mathrm{u}(\br{F}),
		\end{align}
	where we made explicit that $\fgg^\mathrm{u}$ is a function of $\br{F}$. 
	This mean-field model shows clear bistability for
	the same parameter ranges as in simulations (\cref{fig:FGBA_MFFGcg}).
	For the exact same parameters, this mean-field is qualitatively most
	comparable to simulations, but the saddle is much flatter
	(\cref{fig:compsim}, solid blue line versus black dots).
	
	\item Using \emph{percolation theory}. Alternatively, we can estimate forest loss using mean 
	burning probabilities from percolation due to a single fire. The forest perimeter
	exposed to fire for a given realisation is the interface of burnt grass with forest at the end of the fire $\br{FA}^*$. When we take the expectation (for given total grass cover) and forest cells are assumed to be uniformly randomly placed, we have
	\begin{equation}
		\E{\br{FA}^*}=\E{4\br{F}\br{A}^*}=4\br{F}\br{G}\E{Q_{p_\mathrm{g}}},
	\end{equation}
	where $\E{Q_{p_\mathrm{g}}}$ is the mean proportion of grass that burns for given $p_\mathrm{g}$ and 
	$\br{G}$ (see \cref{eq:Qdef}; shown in \cref{fig:percolationshort}b). 
	The loss per fire in this case is then
	\begin{equation}\label{eq:losspfpu2}
		\Delta \br{F}_\mathrm{pu}=4p_\mathrm{f}\br{F}\br{G}\E{Q_{p_\mathrm{g}}},
	\end{equation}
	such that the loss function is (multiplying by $\phi N\br{G}$)
	\begin{equation}\label{eq:losspu2}
		\Delta_{\mathrm{F},\mathrm{pu}}^{\mathrm{loss}}=4p_\mathrm{f}\phi N\br{G}^2\br{F}\E{Q_{p_\mathrm{g}}}. 
	\end{equation}
	The final mean-field model is then
		\begin{align}\label{eq:MF2t_Qpg}
			\ddt\br{F} &=\beta(1-\br{F})+4\alpha\br{F}(1-\br{F})-\gamma\br{F}-4p_\mathrm{f}\phi N(1-\br{F})^2\br{F}\E{Q_{p_\mathrm{g}}}.
		\end{align}
	This mean-field model has very similar steady states as \cref{eq:MF2t_FGcg} but the
	saddle is slightly lower (dotted blue line in \cref{fig:compsim}).
	In the limit of large domain size, $\E{Q_{p_\mathrm{g}}}$ may be replaced by the percolation probability
	$P_{\infty}$, which is defined as the probability that a grass cell belongs to the giant
	component \cite{Christensen2005} (see also \cref{fig:percolation}a).
\end{enumerate}

Both of the estimates above assume that $p_\mathrm{f}=0$ for fire spread and that $p_\mathrm{f}$ 
is small for loss of (uniformly randomly placed) forest due to fire. The first further assumes that fire spreads perfectly on grass ($p_\mathrm{g}=1$).
Therefore the two estimates are equivalent when the spreading process is pure site percolation,
for which $\fgg=4\E{Q_1}\br{F}\br{G}$ (equating \cref{eq:losspu1} and \cref{eq:losspu2}), which 
is confirmed by \cref{fig:lossperfire}b (`$\times$' and `$+$' symbols).
Comparing the estimates to recorded forest loss in fire simulations where only the assumption 
of random placement is
taken (`$\circ$' in \cref{fig:lossperfire}b), one sees that the second estimate (`$\cdot$'
in \cref{fig:lossperfire}b) is more
accurate than the first estimate (`$\times$' in \cref{fig:lossperfire}b), despite that it carries more assumptions. Hence, the error due to the 
assumption of
grass perfectly spreading compensates the error by the assumption of forest perfectly blocking fire.
We expect that the difference between the two approaches
will be smaller for landscapes with spatial aggregation of forest, where
fires spread in pockets of high grass cover, 
for which $\E{Q_1}-\E{Q_{0.9}}$ is smaller (\cref{fig:lossperfire}a). 

\subsection{Detailed comparison against simulations} 
Here, we compare the time-separated mean-field models to the simulations 
of the main text and also to other simulations with
different spreading ranges for forest and/or fire.
In \cref{fig:compsim}, dots with error bars are simulations
and lines are mean-field approximations. The black dots are
the simulations from the main text, with nearest-neighbour
spreading for both fire and forest.
Purple dots are simulations where both fire and forest
can spread to any other cell. Blue dots are simulations
where forest can spread to any other cell, but fire
spreads along nearest neighbours.
And finally, orange dots denote simulations where
forest spread occurs in a Gaussian neighbourhood
with standard deviation 60m (two cells).

All simulations and approximations agree fairly well on the parameter 
value where the lower saddle-node bifurcation occurs.
All except the fully well-mixed case agree on the 
stable steady states. The disagreement occurs particularly
for the unstable steady states, where the effect 
of spatial structure is hence most pronounced
and process-dependent.
There is little or no bistability in the well-mixed two-timescale
mean-field (purple line) but its good agreement with
uniformly mixed simulations (purple dots) further
indicates the validity of the assumption of time-scale separation 
with well-separated fire events.
The mean field where fire is a site percolation process
on landscapes with uniform random placement of forest
(blue line), is qualitatively more
correct but it has a much flatter saddle than the simulations
(black dots). 
Hence, even the mean-field model that
takes into account the effects of percolation while keeping forest
cells spatially uncorrelated remains strongly biased due to the
importance of spatial aggregation of forest cells.
Taking larger neighbourhood sizes in the simulations does
not change this (orange dots). 
Even compared to simulations
with uniform forest dispersal (blue dots), the
mean field with site percolation shows
some bias, indicating that the fire spreading alone already induces
some spatial structure. This is most likely caused by lower survival rates
of solitary compared to aggregated forest cells.

The bias of the mean field is even more apparent in the dynamics. 
\Cref{fig:sim3phivsMF_deco} shows the same
figure as \cref{fig:scatterplots_changes}, but with the corresponding
values of the mean field with site percolation on top.
Panels a--c show large differences between $\br{FG}$
and $\fgg$ of simulations (scattered dots) versus those from the 
mean field (curves). In particular, while for the mean
field, the perimeter is the parabola 
$\br{FG}=4\br{F}\br{G}=4\br{F}(1-\br{F})$,
the perimeter of simulations lies below this parabola for any 
$\br{F}$. That the simulated perimeter is lower
for given forest area means that forest is
more spatially aggregated in simulations. This results in
lower forest growth rate at any cover value (panels d--f)
because fewer forest cells can expand into grass. 
Fire-induced damage is lower below $\br{F}\approx0.4$ and higher above
(panels d--f). 
This is so because damage per fire (at given cover)
is determined by two effects: exposure of forest and clustering of grass.
Below $\br{F}\approx0.4$, there is no clustering, such that only decreased
exposure due to aggregation can decrease forest loss. Above $\br{F}\approx0.4$,
aggregation decreases clustering, such that
grassland stays fully connected at higher
forest cover than in the case with uniform random placement,
with larger fires as a consequence. This further leads to
an upward shift of the unstable forest state compared to the mean field 
(panels (g--i), see also \cref{fig:compsim}).
The effect of forest aggregation on fire spread
has an equivalent in disease spread: in the SIR process,
aggregation of immune individuals lowers the epidemic threshold,
such that it elevates the population immunisation threshold to eradicate the epidemic 
\cite{Keeling1999}. Note though, that, as argued above, the equivalent epidemic process
to tropical fire spread in forest-grassland landscapes is not the regular SIR process, 
but one that has a mix of two 
populations: susceptibles (grass) and imperfectly immunised individuals (forest).

\section{Evolution of fronts --- heterogeneous states}\label{sec:supp:fronts}

Here, we illustrate the case where grass and forest are initially separated into
two contiguous areas with their interface extending along a straight line.
Because for this type of initial conditions, the single-cluster approximation 
(\cref{eq:dFdtclust1ss}) is valid, we can focus on the evolution of the interface.
As spontaneous conversion between forest and grass (with rates $\beta$ and $\gamma$) increases independence between cells and promotes homogeneity at large scale, we expect the effects of heterogeneous initial conditions to be most persistent when the spontaneous conversion rates $\beta$ and $\gamma$ are small.
Therefore, we will set $\beta=\gamma=0$, for which
\cref{eq:dFdtclust} becomes
	\begin{equation}\label{eq:dFdt:nospont}
		\frac{\mathrm{d}\br{F}}{\mathrm{d}t} = (\alpha - \phi p_\mathrm{f} N \br{G}) \br{FG}.
	\end{equation}
Hence, the precise shape of the interface $\br{FG}$ does not affect the location of the steady states, only
the rate at which they are approached or receded from. 
The trivial steady states of \cref{eq:dFdt:nospont} are $\br{F}=0$ and $\br{F}=1$ (where $\br{FG}=0$), 
which are stable, and between them, there is the saddle
	\begin{equation}\label{eq:dFdt:nospont:saddle}
		\br{F}^*=1-\frac{\alpha}{\phi N p_\mathrm{f}}.
	\end{equation}
As seen in \cref{fig:linearinterface_analysis}, this analytical prediction (solid black) 
matches the 
controlled simulations with $p_\mathrm{g}=0.9999$ (shaded blue). For $p_\mathrm{g}=0.9$ (shaded red), which
we used before, there is a small bias. 
In the limit of $N\to\infty$, \cref{eq:dFdt:nospont:saddle} converges to $\br{F}^*_\infty=1$, 
implying that in an infinite domain, any positive fire rate leads to extinction of forest 
below $\br{F}^*=1$. When ignoring the effect of ash, this would also occur for heterogeneous 
initial conditions. That is, considering an infinite domain with many grass clusters of which the size is a random
variable (with support $[0,\infty)$), there will be initial grass clusters of arbitrarily large size, which will expand and eventually drive forest extinct.
However, such determinism does not occur in
the simulations because at high fire rates, patches with ash start to block fires,
and the rate of exposure of the forest interface to fire
becomes limited by the rate at which ash is converted back into grass.
As a first correction for this, one can multiply $p_\mathrm{f}$ with the average proportion of 
grass sites that are in the ash state after the expected waiting time between fires
$1-\exp(-\lambda/\phi N)$, such that 
$\br{F}^*_\infty=1-\alpha/\lambda p_\mathrm{f}$. 
Keeping in mind that we are focusing on heterogeneous states, 
the analysis here implies that for $\gamma=\beta=0$, there is a critical 
patch size above which the forest patch expands
and below which it contracts. The intuition is that above the critical forest patch size,
there is not enough grass area to reach the minimum number of ignitions
required to erode the forest.

\begin{figure}[h]
	\begin{minipage}[c]{0.65\linewidth}
		\includegraphics[width=1\linewidth]{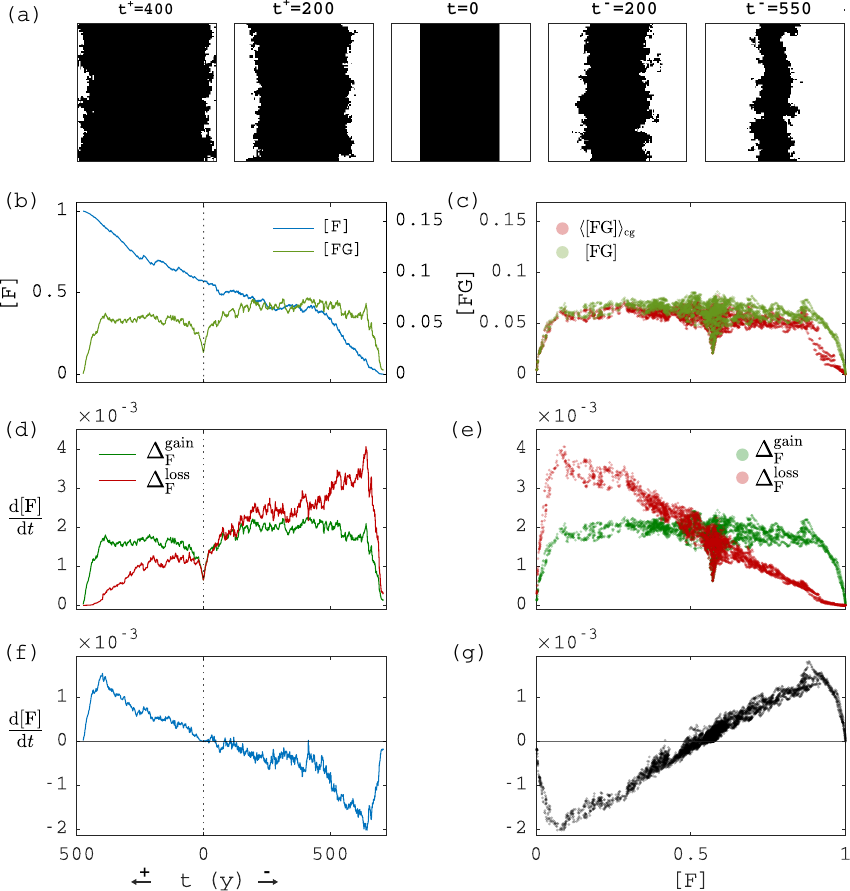}
	\end{minipage}\hfill
	\begin{minipage}[b]{.3\linewidth}
		\caption{Perimeter quantities and change rates according to \cref{eq:dFdtclust} when initial conditions are heterogeneous and there are no spontaneous transitions ($\beta{=}\gamma{=}0$) for $\phi{=}6.98\cdot10^{-5}$, displayed as in \cref{fig:gainlossterms,fig:scatterplots_changes}. (a) Snapshots of the domain at indicated times. (b,d,f) Cover, interface and rate of change from \cref{eq:dFdtclust} versus time. (c,e,g) Interface, grass-cluster weighted average of the interface, and rate of change from \cref{eq:dFdtclust} versus cover. Remaining parameters are $\alpha{=}0.03,\rho_\mathrm{g}{=}10^{10},\rho_\mathrm{f}{=}1.11\cdot10^{5},\mu{=}10^6,\lambda{=}5$.}
		\label{fig:linearinterface}
	\end{minipage}
\end{figure}

In \cref{fig:linearinterface}, we show how the dynamics and steady states
arise from $\br{FG}$, as we did in \cref{fig:gainlossterms} and \cref{fig:scatterplots_changes}, but now starting with separated patches of
grass and forest that interface on a line on both sides (showing for $\br{F}_0=0.57$).
From $\br{FG}(0)=2L/N=2N^{-1/2}=0.02$, the interface
quickly gains roughness due to the dynamics, until it reaches a
steady state of $\br{FG}^*\approx6L/N=6N^{-1/2}=0.06$ (see \cref{fig:linearinterface}a--c).
\Cref{fig:linearinterface}b
confirms that $\fgg\approx\br{FG}$ (except when
forest cover approaches $\br{F}=0$ or $\br{F}=1$), confirming that the single-cluster approximation is valid.
Away from $\br{F}=0$ and $\br{F}=1$, $\br{FG}$ stays about constant as the forest interface moves
(\cref{fig:linearinterface}b--c). 
Therefore the gain and loss terms (as defined in \cref{eq:slow:gain,eq:slow:loss2}) are now, respectively,
constant and linearly decreasing with $\br{F}$ (\cref{fig:linearinterface}e),
such that $\mathrm{d}\br{F}/\mathrm{d}t$ increases linearly with $\br{F}$ (\cref{fig:linearinterface}g), 
except near $\br{F}=0$ and $\br{F}=1$, where it connects to 0, as here $\br{FG}=0$.

\section{A note on finite sizes and bistability} \label{sec:finite}

In simple bistable chemical systems, it is known 
that bistability converges to an abrupt phase transition
in the thermodynamic limit ($N{\to}\infty$) \cite{Ge2009,Thiele2019}, at a value known as the Maxwell point \cite[e.g.][]{Goldenfeld2018}, making the macroscopic state of the system
deterministically dependent on the parameters (e.g. pressure or temperature). With only 
forest and grass or provided that savanna and forest components are
sufficiently decoupled, the same behaviour occurs in spatial mean-field
models of tropical forest, with a front between forest and nonforest that is
depends deterministically on environmental drivers \cite{Wuyts2019}.
We do not expect such determinism as $N{\to}\infty$ to arise in the FGBA cellular 
automaton. 
The infinite FGBA cellular automaton possesses grass clusters of arbitrary size,
such that, even when assuming that fire spreads instantly on grass, there will 
always be some parts of the forest perimeter shielded from intruding fires by
adjacent ash cells. Were it not for this shielding effect, then there would be
a deterministic dependence of the dynamics on fire ignition rate away from the absorbing states: $\phi{=}0$ would lead to forest spread while $\phi{>}0$ would lead 
to forest extinction 
(see \cref{sec:supp:fronts}, 
for $\beta{=}\gamma{=}0$). 
In reality, finite fire spreading rates and, in particular, 
the effects of heterogeneity in space or time
impose stronger limits on the reach of fires.

In finite
domains,
both the cellular automaton and scalar reaction-diffusion equations (with bistable reaction term) show bistability 
due to critical patch
sizes or domain shapes, and dependence on interfacial characteristics \cite[e.g.][]{Liehr2013,Goel2020,Levin1979}, but this correspondence requires further scrutiny. 
In realistic scenarios, we then suspect that the amount of bistability depends (besides the parameters) on the ratio
of the characteristic interaction scale and the scale of observation.
The range of interaction in turn depends on e.g. fire spreading and/or plant dispersal ranges. 
E.g., if we take as interaction scale the typical size of a
savanna fire (assuming that it exists) ${\cal O}(10^{1\ldots2}\mathrm{km^2})$ \cite{Andela2019}, 
this corresponds to an area in the cellular automaton of $100{\times}100$ to $300{\times}300$ cells. 
This also corresponds to our observation scale (domain size) in the main text. Hence, it may be
that the bistability observed in our work is a finite-size effect, and that a larger observation
scale leads to more gradual transitions due to existence of multiple stable patterns 
\cite{Rietkerk2021,Bastiaansen2022}.


\section{How to include fire parameters in existing mean-field models}\label{sec:inclfireparams}
\label{sec:fireinclmf} 
Previously derived mean-field models of tropical tree cover bistability did not include parameters
that relate to fire. Here we give suggestions on how to include fire ignition rate and
the appropriate percolation quantities, focusing on the Staver-Levin mean-field
model \cite{Schertzer2014,Staver2012} of tropical savanna and/or forest bistability. We assume the reader
is familiar with \cite{Schertzer2014,Staver2012}.

\begin{figure}[H]
	\centering
	\includegraphics[scale=.69]{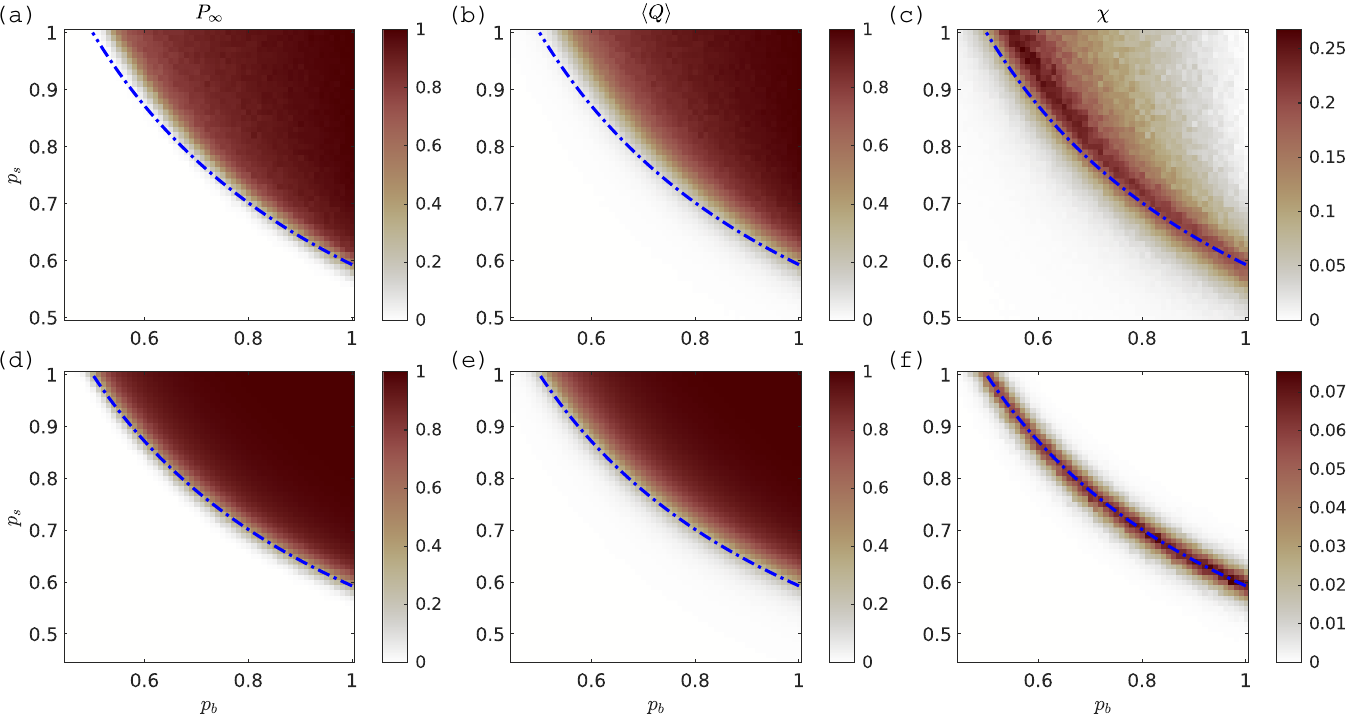}
	\caption{GBA process with $\phi{=}\lambda{=}0$ (equivalent to SIR spreading on the square lattice) in terms of bond occupation probability $p_b:=p_\mathrm{g}$ (\cref{eq:occupation:prob}) and site occupation probability $p_s:=\br{G}_0$ (a-c) and compared to standard mixed site-bond percolation (d-f). Shown quantities as a function of bond and site occupation probability (based on 1024 realisations for (a-c) and on 512 realisations for (d-f)): (a,d) percolation probability $P_\infty$, (b,e) mean proportion of burnt grass cells $\E{Q}$ (see \cref{eq:Qdef}), and, (c,f) susceptibility $\chi{=}\E{Q^2}/\E{Q}$. Percolation	probability is defined as the probability that any grass cell belongs to the giant component \citep{Christensen2005}. Susceptibility is defined here as in \cite{Hebert-Dufresne2019}, using $Q$ as order parameter. The dash-dotted blue line indicates the location of the infinite-size percolation threshold for uncorrelated mixed site-bond percolation (taken from \citep{Tarasevich1999}). For a domain size of 100x100 cells and at the shown resolution, the percolation threshold of mixed site-bond percolation matches that of the infinite size system (f). The GBA process' percolation threshold lies at higher values (c) due to spatial correlation of $p_\mathrm{g}$ as explained in the text. The top row is more noisy than the bottom row because for standard mixed percolation, we were able to obtain the whole distribution of cluster sizes for a each realisation	and computed their statistics using percolation theory \citep{Christensen2005}, whereas for the GBA process, each simulation only resulted in one sample. The colour scale was taken from \cite{Crameri2018}.}\label{fig:percolation}
\end{figure}

By running an infection process on clusters obtained by standard site percolation, 
\cite{Schertzer2014} used a mixed
site-uncorrelated bond-correlated percolation process for fire (although not using this terminology).
The site percolation is due to uniformly randomly distributed tree cells perfectly blocking fires and the 
bond percolation due to flammable cells (grass and savanna saplings in \cite{Schertzer2014}) spreading fires
with a given probability. The correlation in bond occupation probability occurs due to the infection
dynamics, as explained in \cref{sec:percsingle}.
One can use the complement of the burning probability of this mixed percolation process, 
i.e. $1-\E{Q_{p_\mathrm{b}}(p_s)}$, as survival 
probability instead of that used in \cite{Schertzer2014}
(see \cref{fig:percolation}b for a plot of $\E{Q_{p_\mathrm{b}}(p_s)}$ as a function of the infection probability between flammable cells $p_b$ and
the probability of a cell being flammable $p_s$). If one does so, 
one can write the mean-field recruitment 
rate of savanna saplings during a small time interval as $\omega(p_b,p_s)\br{S}$, where $\omega$ is
\begin{equation}
	\omega(p_b,p_s):=\max[\omega_0-\phi N p_s \E{Q_{p_\mathrm{b}}(p_s)}\Delta\omega,0],
\end{equation}
with $\Delta\omega{>}0$ the per fire decrease in recruitment rate due to burning
and $\phi$ the fire ignition rate in flammable cells. Note that, according to
\cite{Schertzer2014}, the total flammable area is $p_s=\br{S}+\br{G}$ (i.e., the 
area of grass and savanna saplings). The reasoning is that there are on average $\phi N p_s$ ignitions,
each causing a fire that on average affects a proportion $\E{Q_{p_\mathrm{b}}(p_s)}$ of
flammable cells, thereby lowering the recruitment rate by an amount $\Delta\omega$.
If $\br{S}$ and $\br{T}$ cells are uniformly randomly placed in the
area affected by fire, then it follows that the recruitment rate is $\omega(p_b,p_s)\br{S}$.

For the approximate effect on forest trees \cite[as included in][]{Staver2012}, one
needs to take into account that forest trees are assumed \cite[in][]{Staver2012} 
to block fires perfectly. Therefore, they are not in the flammable part of the landscape,
but instead share an interface with it. The mean-field rate of forest loss due to fire
during a small time interval is then $\zeta(p_b,p_s)\br{F}$, where $\zeta$ is
\begin{equation}
	\zeta(p_b,p_s):=4\phi N p_s^2 p_\mathrm{f} \E{Q_{p_\mathrm{b}}(p_s)},
\end{equation}
with $p_s=\br{S}+\br{G}$ also. The reasoning here is as follows. There are on average $\phi N p_s$ ignitions,
each causing a fire that affects a proportion $\E{Q_{p_\mathrm{b}}(p_s)}$ of the landscape. Assuming that occurrences of 
burnt and forest cells are uncorrelated, one can write the interface between them
as the number of forest-burnt pairs in a lattice: $4 (\E{Q_{p_\mathrm{b}}(p_s)}p_s) \br{F}$. For each such pair, there is a probability $p_\mathrm{f}$ of spreading into forest, such that (when using the approximation of
small $p_\mathrm{f}$ as in the main text), the resulting forest loss is $4\phi N p_s^2 p_\mathrm{f} \E{Q_{p_\mathrm{b}}(p_s)}\br{F}$.

For large domains, one may replace $\E{Q_{p_\mathrm{b}}(p_s)}$ by the percolation probability $P_\infty(p_b,p_s)$ (shown in \cref{fig:percolation}a), 
which is the probability that a flammable cell is part of the giant connected component.

\newpage
\section{Additional figures}

\noindent
\begin{minipage}{.55\columnwidth}
	\raggedleft
\end{minipage}
\hfill
\begin{minipage}{.4\columnwidth}
	\raggedright
	\begin{figure}[H]
		\begin{center}
			\includegraphics[width=.77\columnwidth]{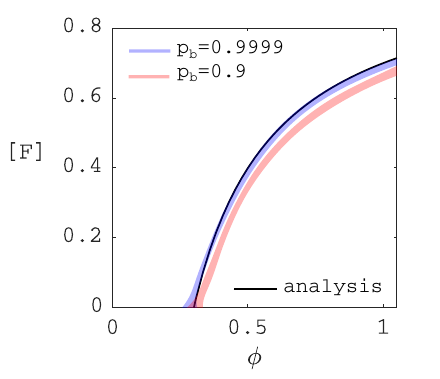}
			\caption{Saddle of \cref{eq:dFdt:nospont} via \cref{eq:dFdt:nospont:saddle} (single cluster -- solid black) compared to controlled simulations (shaded red: $p_\mathrm{g}=\rho_\mathrm{g}/(\rho_\mathrm{g}+\mu)=0.9$, shaded blue: $p_\mathrm{g}=0.9999$) in case of heterogeneous initial conditions and without spontaneous interactions ($\beta{=}\gamma{=}0$). Other parameters: $\alpha{=}0.03,\rho_\mathrm{f}{=}1.11\cdot10^{5},\mu{=}10^6,\lambda{=}5$.}
			\label{fig:linearinterface_analysis}
		\end{center}
	\end{figure}
\end{minipage}
\noindent

\end{document}